\documentclass[fleqn,usenatbib]{mnras}

\usepackage{newtxtext,newtxmath}

\usepackage[T1]{fontenc}

\DeclareRobustCommand{\VAN}[3]{#2}
\let\VANthebibliography\thebibliography
\def\thebibliography{\DeclareRobustCommand{\VAN}[3]{##3}\VANthebibliography}

\usepackage{graphicx}	
\usepackage{amsmath}	

\usepackage{amssymb}	
\usepackage{hyperref}
\usepackage{multirow}
\usepackage{rotating}
\usepackage{subfloat}
\usepackage{xcolor}
\usepackage{longtable, makecell}

\usepackage{siunitx}
\usepackage{tablefootnote}
\usepackage{footnote}
\usepackage{float}
\usepackage{placeins}
\usepackage{afterpage}
\usepackage{perpage}
\usepackage{pdflscape}

\makeatletter
\newlength{\abovecaptionskip}
\makeatother

\usepackage{threeparttablex}
\usepackage{threeparttable}
\usepackage{xspace}
\usepackage{tabularx}
\usepackage{orcidlink}
\usepackage{threeparttable}
\usepackage{url}

\newcommand{\hii}{\mbox{H\,\textsc{ii}}\xspace}
\newcommand{\msun}{\mbox{M$_\odot$}\xspace}
\newcommand{\Lsun}{\mbox{L$_\odot$}\xspace}
\newcommand{\kms}{\mbox{km\,s$^{-1}$}\xspace}

\newcommand{\hsco}{\mbox{H$^{13}$CO$^+$}\xspace}

\newcommand{\hscn}{\mbox{H$^{13}$CN}\xspace}
\newcommand{\hcsn}{\mbox{HC$_{3}$N}\xspace}

\newcolumntype{N}{S[table-format=2.2]}

\newcommand{\numred}{6\xspace}
\newcommand{\numblue}{10\xspace}
\newcommand{\numtot}{16\xspace}
\newcommand{\bluevrange}{[-92, -54]\,km\,s$^{-1}$\xspace}
\newcommand{\redvrange}{[-37, 13]\,km\,s$^{-1}$\xspace}
\newcommand{\MassMsun}{2.0}
\newcommand{\MomentumMsunkms}{62.3}
\newcommand{\EnergyErg}{$2.5\times10^{46}$}
\newcommand{\tdyn}{5900}

\title[Explosive Outflow Candidate in I16119]{ALMA Reveals an Explosive Outflow Candidate in IRAS\,16119--5048}

\author[Luo et al.]{
Yongquan Luo\textsuperscript{\orcidlink{0009-0000-8349-7355}},$^{1,2}$\thanks{E-mail: luoyongquan@xao.ac.cn}
Jianjun Zhou\textsuperscript{\orcidlink{0000-0003-0356-818X}},$^{1,2,3}$\thanks{E-mail: zhoujj@xao.ac.cn}
Tie Liu\textsuperscript{\orcidlink{0000-0002-5286-2564}},$^{4,2}$\thanks{E-mail: liutie@shao.ac.cn}
Jarken Esimbek\textsuperscript{\orcidlink{0000-0003-4910-1390}},$^{1,2,3}$
Siju Zhang\textsuperscript{\orcidlink{0000-0002-9836-0279}},$^{5}$
\newauthor
Xindi Tang\textsuperscript{\orcidlink{0000-0002-4154-4309}},$^{1,2,3}$
Sami Dib\textsuperscript{\orcidlink{0000-0002-8697-9808}},$^{6}$
Prasanta Gorai\textsuperscript{\orcidlink{0000-0003-1602-6849}},$^{7,8,9}$
Mika Juvela\textsuperscript{\orcidlink{0000-0002-5809-4834}},$^{10}$
Leonardo Bronfman\textsuperscript{\orcidlink{0000-0002-9574-8454}},$^{5}$
\newauthor
Patricio Sanhueza\textsuperscript{\orcidlink{0000-0002-7125-7685}},$^{11}$
Jihye Hwang\textsuperscript{\orcidlink{0000-0001-7866-2686}},$^{12,13}$
Fengwei Xu\textsuperscript{\orcidlink{0000-0001-5950-1932}},$^{6}$
Kee-Tae Kim\textsuperscript{\orcidlink{0000-0003-2412-7092}},$^{14,15}$
Guido Garay\textsuperscript{\orcidlink{0000-0003-1649-7958}},$^{5,16}$
\newauthor
Chang Won Lee\textsuperscript{\orcidlink{0000-0002-3179-6334}},$^{14,15}$
Tapas Baug\textsuperscript{\orcidlink{0000-0003-0295-6586}},$^{17}$
L. Viktor Tóth\textsuperscript{\orcidlink{0000-0002-5310-4212}},$^{18,19}$
Gang Wu\textsuperscript{\orcidlink{0000-0003-0933-7112}},$^{1,2,3}$
Dalei Li\textsuperscript{\orcidlink{0000-0001-5494-6238}},$^{1,2,3}$
Yuxin He\textsuperscript{\orcidlink{0000-0002-8760-8988}},$^{1,2,3}$
\newauthor
Yingxiu Ma\textsuperscript{\orcidlink{0000-0002-0776-0753}},$^{1,3}$
Dongdong Zhou,$^{1,3}$
Toktarkhan Komesh\textsuperscript{\orcidlink{0000-0002-3415-4636}},$^{1,20,21}$
Weiguang Ji,$^{1,3}$
Dezhao Meng\textsuperscript{\orcidlink{0009-0000-5764-8527}},$^{1,2}$
\newauthor
Jiasheng Li\textsuperscript{\orcidlink{0009-0008-8664-5681}}$^{1,2}$
\\
Affiliations are listed at the end of the paper
}

\date{Received XXX; accepted XXX}

\pubyear{\the\year{}}

\begin{document}
\label{firstpage}
\pagerange{\pageref{firstpage}--\pageref{lastpage}}
\maketitle

\begin{abstract}

We present a multiwavelength study of the massive star formation region IRAS~16119--5048 (I16119) using ALMA ATOMS Band~3 and QUARKS Band~6 observations, complemented by archival ATCA radio continuum and \textit{Spitzer} mid-infrared data. 
The CO\,(2--1) emission reveals a system of high-velocity streamer-like structures around the central region. 
Using a dendrogram analysis of velocity-channel maps followed by linking in position--position--velocity space, we identify \numtot streamers that are approximately radially distributed and whose projected trajectories converge toward a common central region.
We found that the kinetic energy of the outflows is at least an order of magnitude lower than those of most known explosive outflows, but the mass entrainment rate and momentum rate are high compared with typical protostellar outflows, suggesting that I16119 may represent a low-energy explosive outflow candidate.  
Dense-gas and photodissociation-region tracers reveal shell-like structures associated with the 8\,$\mu$m emission, indicating that feedback from the \hii\ region may influence the streamer morphology. 
The 1.3\,mm continuum emission resolves 27 dense cores along a fragmented filamentary structure. 
Their separations are consistent with thermal Jeans or thermal cylindrical fragmentation, while the collect and collapse scenario is unsupported. 
The dense cores also show evidence of mass segregation, with the most massive cores concentrated near the inferred explosive center. 
We therefore suggest that I16119 is a plausible low-energy explosive outflow candidate, possibly triggered by dynamical interactions among centrally concentrated massive cores.
However, the complex velocity structure and possible contamination from individual core-driven outflows prevent a definitive classification. More sensitive, higher-angular-resolution, and dedicated observations are required to confirm the nature of the outflow in this region.

\end{abstract}

\begin{keywords}
  stars: formation --
  ISM: jets and outflows --
  ISM: \hii region --
  ISM: individual objects: I16119--5048
\end{keywords}

\section{INTRODUCTION}

Explosive outflows represent a rare and highly energetic mode of molecular gas ejection in massive star formation regions. 
They differ from classical bipolar outflows, which are generally interpreted as the consequence of quasi-steady accretion through circumstellar discs, in that they appear to be impulsive events produced by a sudden release of energy. 
Several mechanisms have been proposed, including the dynamical decay of non-hierarchical multiple systems, close stellar encounters, and protostellar mergers, in which gravitational binding energy can be rapidly converted into kinetic energy of the surrounding gas \citep[e.g.,][]{Bally2005, Zapata2009, Bally2011, Rivilla2014, Bally2016, Bally2017, Zapata2019, LiDL2020, ZhaoX2024}. 
Observationally, explosive outflows are characterized by multiple narrow molecular streamers that emerge from a common center, often with red- and blueshifted components overlapping in projection, and by a Hubble--Lema\^itre-like velocity field in which the line-of-sight velocity increases approximately linearly with projected distance from the origin \citep{Zapata2017, Zapata2019, Zapata2020, Zapata2023}.

The best-studied example is Orion BN/KL, where more than one hundred CO streamers, together with shock-excited H$_2$ and [\ion{Fe}{ii}] fingers, trace an energetic event that likely occurred a few hundred years ago and was associated with the dynamical ejection of massive young stars \citep{Bally2017}. 
A small number of additional explosive outflow candidates have since been identified, including DR\,21, G5.89$-$0.39, IRAS\,16076$-$5134, Sh2$-$106, IRAS\,12326$-$6245, G34.26+0.15, and IRAS\,15520$-$5234 \citep{Zapata2013, Guzman2022, Bally2022, Zapata2023, Issac2025, Hoque2026}. 
These systems suggest that explosive events may be linked to the early dynamical evolution of compact massive protoclusters. 
In several cases, the molecular streamers are accompanied by shocked-gas tracers such as SiO, shell-like or compact \hii\ regions, and in some sources quasi-radial magnetic-field morphologies, indicating that the explosive ejecta may interact strongly with the surrounding dense and ionized gas \citep{Cortes2021, FernandezLopez2021, Zapata2023, Guzman2024, Issac2025}.

IRAS\,16119$-$5048, hereafter I16119, is a massive star formation complex located at a distance of 3.42\,kpc, with a systemic velocity of $V_{\rm lsr}=-48.6$\,\kms \citep{LiuXC2024}. 
The associated ATLASGAL clump has a mass of $10^{3.2}$\,\msun, a bolometric luminosity of $10^{4.3}$\,\Lsun, a radius of 0.93\,pc, and a dust temperature of 24\,K \citep{Urquhart2018, LiuTie2020}. 
The region is associated with a cometary \hii\ region \citep{Urquhart2007, Lumsden2013}, and a recent study has revealed hot-core activity traced by CH$_3$CN emission \citep{Meng2026}. 
These properties indicate that I16119 is an active massive protocluster in which feedback from both young stellar objects and ionized gas may influence the surrounding molecular material.

In this work, we present a multiwavelength study of I16119 using high-angular-resolution ALMA observations from the ATOMS and QUARKS surveys, complemented by archival ATCA radio continuum and \textit{Spitzer} mid-infrared data.  
We identify multiple CO\,(2--1) streamer-like structures emerging from the central region, examine their morphology and position--velocity properties, and estimate their mass, momentum, kinetic energy, and dynamical time-scale. We further compare these properties with those of known explosive outflows and more typical protostellar outflows, in order to assess whether I16119 represents a new, relatively low-energy candidate member of the explosive-outflow class. In addition, we use SiO\,(5--4), dense-gas tracers, H40$\alpha$ emission, and the 1.3\,mm continuum to investigate the interaction between the candidate explosive outflow, the adjacent cometary \hii\ region, and the compact population of dense cores. This allows us to explore whether ionized-gas feedback, fragmentation, mass segregation, and dynamical interactions among the central dense cores may have contributed to the observed outflow phenomenology.

This paper is organized as follows. Section~\ref{sect:observations} describes the ALMA, ATCA, and infrared data used in this work. In Section~\ref{sect:results_discussion}, we present the main results and discussion, including the identification of the CO\,(2--1) streamers, the derivation of the outflow physical parameters, the evidence supporting an explosive-outflow interpretation, the interaction between the molecular clump and the adjacent \hii\ region, the properties of the 1.3\,mm dense cores, and the implications of fragmentation and mass segregation for the possible driving mechanism. Finally, Section~\ref{sect:summary} summarises our main conclusions.

\section{OBSERVATIONS}
\label{sect:observations}

\subsection{ALMA observations}
\subsubsection{ATOMS Band~3 observations}
I16119 was observed as part of the ALMA Three-millimeter Observations of Massive Star-forming regions (ATOMS) survey (Project ID: 2019.1.00685.S; PI: Tie Liu), which targeted 146 active massive protocluster regions in Band~3 using both the ALMA 12-m array and the 7-m Atacama Compact Array \citep{LiuTie2020}.

The ACA observations of I16119 were conducted on 2019 November 17--18, with projected baselines ranging from 8.9 to 48.0\,m. The 12-m-array observations were conducted on 2019 November 4, with projected baselines ranging from 15.1 to 500.2\,m. The typical on-source integration times of the ATOMS observations were approximately 8 and 3\,min per target for the ACA and 12-m arrays, respectively. The observations were carried out in single-pointing mode using the Band~3 receivers in dual-polarization mode.

The lines used in this work include H$^{13}$CO$^+$\,(1--0), H$^{13}$CN\,(1--0), CCH\,(1--0), HC$_3$N\,(11--10), and H40$\alpha$. The corresponding velocity resolutions are 0.42\,km\,s$^{-1}$ for H$^{13}$CO$^+$\,(1--0), H$^{13}$CN\,(1--0), and CCH\,(1--0), and 2.9\,km\,s$^{-1}$ for HC$_3$N\,(11--10) and H40$\alpha$.

Calibration and imaging were performed using \texttt{CASA} version 5.6 \citep{CASA2022}. For this study, we used the combined ACA and 12-m-array data products to recover both compact and relatively extended emission. The continuum emission was constructed using line-free channels, and primary-beam correction was applied to the final continuum and spectral-line images.

The final 3\,mm continuum image has an angular resolution of $1.58^{\prime\prime}$, an rms sensitivity of 0.13\,mJy\,beam$^{-1}$, and a maximum recoverable scale of approximately $60^{\prime\prime}$. The rms sensitivities of the spectral-line cubes are 0.22, 0.31, 0.22, 0.11, and 0.11\,K for H$^{13}$CO$^+$\,(1--0), H$^{13}$CN\,(1--0), CCH\,(1--0), HC$_3$N\,(11--10), and H40$\alpha$, respectively.

\subsubsection{QUARKS Band~6 observations}
I16119 was also observed as part of the Querying Underlying mechanisms of massive star formation with ALMA-Resolved gas Kinematics and Structures (QUARKS) survey (Project ID: 2021.1.00095.S; PIs: Lei Zhu, Guido Garay, and Tie Liu). The QUARKS survey complements the ATOMS observations with higher-angular-resolution Band~6 observations at approximately 1.3\,mm. The observations were conducted in single-pointing mode using the ACA 7-m array and the ALMA 12-m array in the C-2 and C-5 configurations \citep{LiuXC2024, XuFW2024, YangDT2025}. The typical on-source integration times were approximately 5, 1, and 5\,min per target for the ACA, C-2, and C-5 configurations, respectively.

The observations employed the Band~6 receivers in dual-polarization mode. Four spectral windows were centered at 217.918429, 220.318632, 231.369566, and 233.519748\,GHz. Each spectral window had a bandwidth of 1.875\,GHz for the
12-m-array observations and 2.0\,GHz for the ACA observations. The four spectral windows have native velocity resolutions of 1.344, 1.329, 1.266, and 1.254\,km\,s$^{-1}$, respectively.

The lines used in this study include SiO\,(5--4), CH$_3$OH\,(4$_2$--3$_1$), $^{13}$CO\,(2--1), SO\,(6--5), and CO\,(2--1). SiO\,(5--4) and CH$_3$OH\,(4$_2$--3$_1$) are covered by the first spectral window and therefore have a velocity resolution of 1.344\,km\,s$^{-1}$. The $^{13}$CO\,(2--1) and SO\,(6--5) transitions are covered by the second spectral window and have a velocity resolution of 1.329\,km\,s$^{-1}$. The CO\,(2--1) transition is covered by the third spectral window and has a velocity resolution of 1.266\,km\,s$^{-1}$.

The calibrated ACA, C-2, and C-5 visibility data were combined and imaged using \textsc{CASA} version 6.5. Line-free channels from all four spectral windows were used to construct the 1.3\,mm continuum image. The continuum data were imaged using the multi-scale multi-frequency synthesis algorithm with \texttt{nterms=2} and Briggs weighting with \texttt{robust=0.5} \citep{LiuXC2024}. The spectral-line continuum was subtracted in the visibility domain, and the line cubes were deconvolved using the multi-scale algorithm. Primary-beam correction was applied to all final images.

The resulting 1.3\,mm continuum image has an angular resolution of approximately $0.3^{\prime\prime}$ and an rms sensitivity of approximately 0.12\,mJy\,beam$^{-1}$. The combined observations have a maximum recoverable scale of approximately $30^{\prime\prime}$. The spectral-line cubes have an rms sensitivity of approximately 0.6\,K per velocity channel.

More details on the observational setups of the ATOMS and QUARKS surveys, as well as their data reduction processes, can be found in \citet{LiuTie2020, LiuXC2024, XuFW2024, YangDT2025}.

\subsection{\textbf{Spitzer} infrared data}
\textit{Spitzer} 4.5\,$\mu$m and 8\,$\mu$m data are obtained from the Galactic Legacy Infrared Mid-Plane Survey Extraordinaire (GLIMPSE) \citep{Benjamin2003} with the $5\sigma$ sensitivities of 0.2 and 0.4\,mJy, respectively. The corresponding spatial resolutions range from $1.5''$ to $1.9''$ \citep{Fazio2004}.

\subsection{Australia Telescope Compact Array Data}
We made use of archival radio continuum data obtained with the Australia Telescope Compact Array (ATCA; located at the Paul Wild Observatory, Narrabri, New South Wales), originally presented by \citet{Urquhart2007}. The archival observations were conducted in the 6C and 6D configurations, yielding a maximum baseline of 6\,km. A bandwidth of 128\,MHz was employed at each frequency, centred at 8640\,MHz (3.6\,cm) and 4800\,MHz (6\,cm), with corresponding primary beam sizes of 5\farcm5 and 9\farcm9. The typical synthesised beam sizes are $\sim$\,1\farcs5 at 3.6\,cm and $\sim$\,2\farcs5 at 6\,cm, and the largest well-imaged structures are $\sim$\,20\arcsec\ and $\sim$\,30\,\arcsec, respectively. The theoretical rms sensitivities are 0.21\,mJy\,beam$^{-1}$ (3.6\,cm) and 0.22\,mJy\,beam$^{-1}$ (6\,cm), while the typical achieved rms noise levels are $\sim$\,0.32\,mJy\,beam$^{-1}$ and $\sim$\,0.27\,mJy\,beam$^{-1}$. Images were produced with pixel scales of 0\farcs33 (3.6\,cm) and 0\farcs6 (6\,cm), ensuring adequate sampling of the synthesised beam \citep{Urquhart2007}.

\section{RESULTS AND DISCUSSION}
\label{sect:results_discussion}
\subsection{Identification of outflows}

\begin{figure*}
    \centering
    \includegraphics[width = 0.95\textwidth]{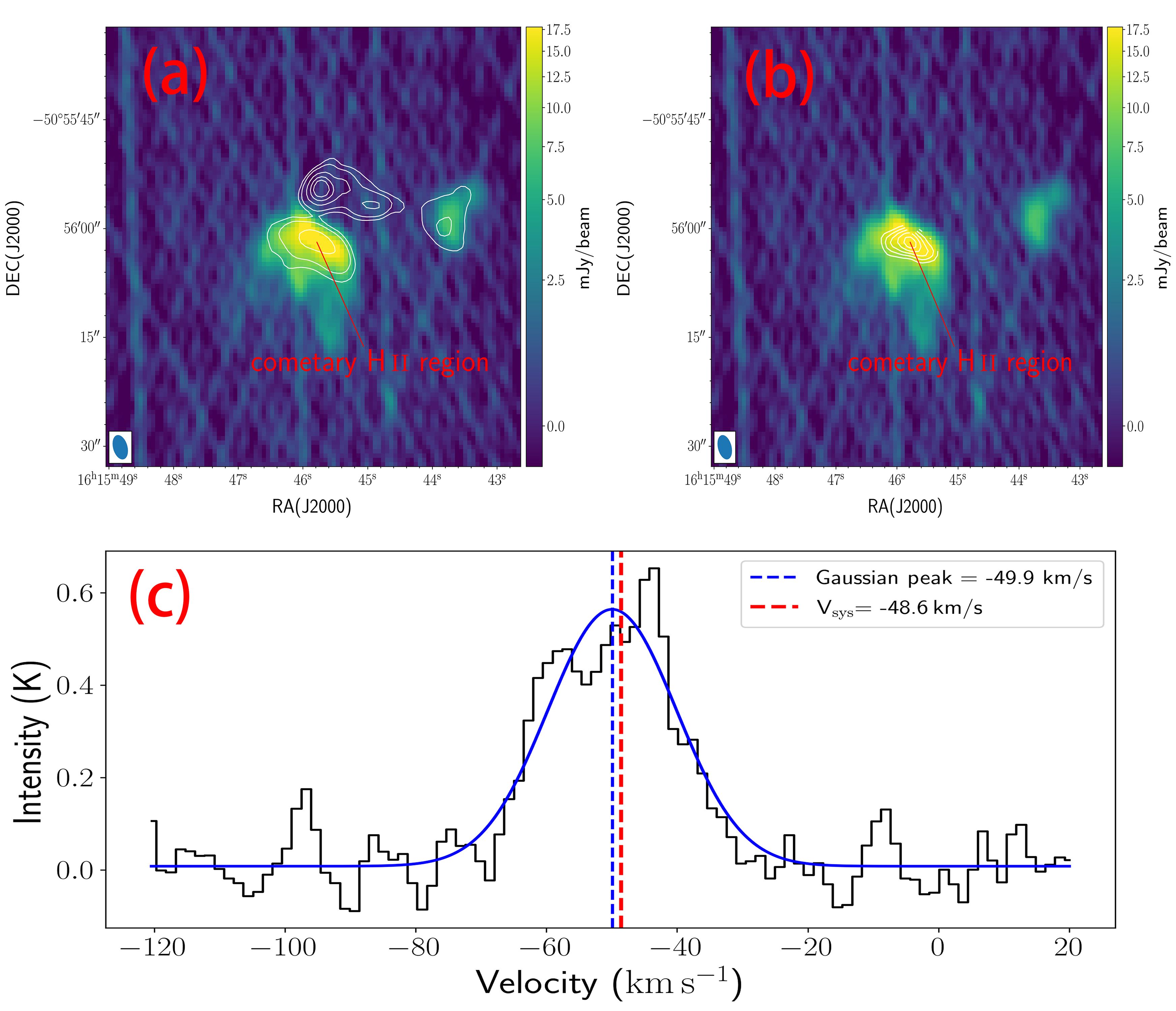}
    \caption{
    Panel (a): ATCA 6\,cm continuum emission toward I16119 shown in color scale.
    White contours show the ALMA 3\,mm continuum emission at
    $[0.1,\,0.15,\,0.3,\,0.5,\,0.7,\,1]\times$ the peak intensity
    (0.024\,mJy\,beam$^{-1}$).
    Panel (b): Same 6\,cm continuum map with white contours showing the integrated
    intensity of the ALMA H40$\alpha$ recombination line integrated from
    $-80$ to $-20$\,km\,s$^{-1}$.
    Contour levels are $[4,5,6,7,8]\sigma$ with
    $\sigma = 0.046$\,Jy\,beam$^{-1}$\,km\,s$^{-1}$.
    Panel (c): H40$\alpha$ spectrum extracted at the peak of the integrated emission.
    The blue curve shows the Gaussian fit.
    The blue dashed line marks the fitted velocity ($-49.9$\,km\,s$^{-1}$),
    and the red dashed line indicates the systemic velocity of I16119
    ($V_{\rm sys}=-48.6$\,km\,s$^{-1}$).
    The synthesized beam of the ATCA 6\,cm continuum is shown in the lower left
    corner of panels (a) and (b).
    }
    \label{fig:H40_alpha}
\end{figure*}

\begin{figure*}
    \centering
    \includegraphics[width = 0.98\textwidth]{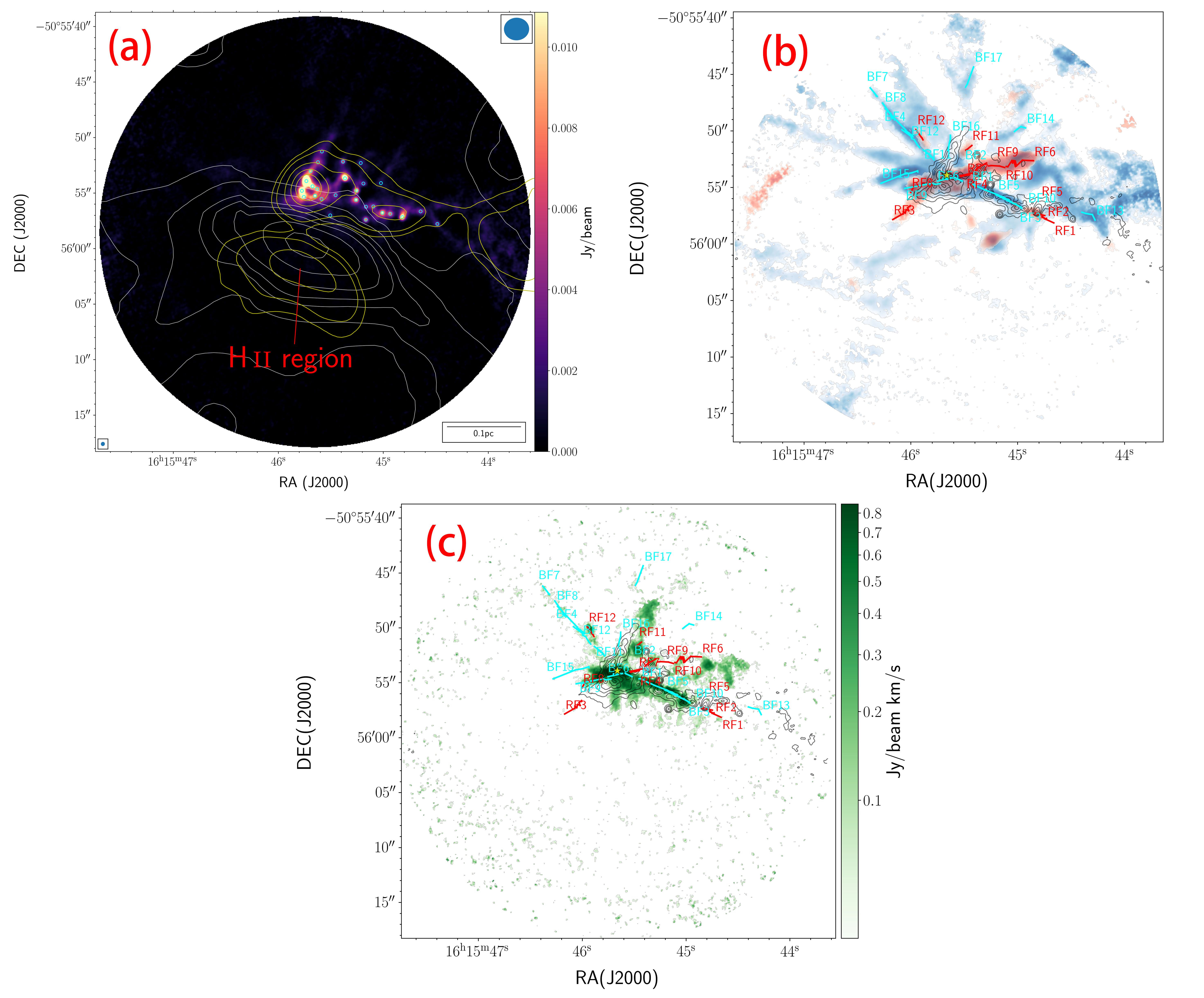}
    \caption{
    Panel (a): 1.3\,mm continuum emission map (color scale), overlaid with 3\,mm continuum emission contours (yellow) and {\it Spitzer} 8\,$\mu$m emission contours (white). 
    The 3\,mm contours are plotted at [0.1, 0.15, 0.3, 0.5, 0.7]$\times$ the peak intensity ($0.024\,\mathrm{Jy\,beam^{-1}}$), while the 8\,$\mu$m contours correspond to [3, 5, 8, 10, 13, 18, 25]$\times\sigma$ ($\sigma=107\,\mathrm{MJy\,sr^{-1}}$). 
    The synthesized beams of the 1.3\,mm and 3\,mm continuum are shown in the bottom-left and top-right corners, respectively. 
    The small cyan circles denote the positions of dense cores identified in Section~\ref{sect:cores_iden}.
    Panel (b): Integrated intensity maps of CO\,(2--1), with the blueshifted ($-92$ to $-57$\,km\,s$^{-1}$) and redshifted ($-37$ to $+13$\,km\,s$^{-1}$) components shown in blue and red, respectively. 
    The identified streamer-like structures are marked with cyan and red lines for the blueshifted and redshifted components, respectively. 
    Panel (c): Integrated intensity map of SiO\,(5--4) emission integrated over $-100$ to $-20$\,\kms, shown in green color scale, with the same streamer-like structure overlays as in panel (b).
    In both panels (b) and (c), black contours show the 1.3\,mm continuum emission at [5, 9, 15, 27, 50, 70]$\times\sigma$ ($\sigma=0.29\,\mathrm{mJy\,beam^{-1}}$), and the yellow star marks the inferred explosion center.
    }
    \label{fig:cont_EO}
\end{figure*}

Panels (a) and (b) of Fig.~\ref{fig:H40_alpha} show the ATCA 6\,cm continuum emission, which exhibits a cometary morphology characteristic of a cometary \hii\ region \citep{Urquhart2007}.
Panel (a) of Fig.~\ref{fig:cont_EO} presents the ALMA 1.3\,mm continuum emission map of I16119. 
The outflow region is located adjacent to a shell-like structure traced by the 8\,$\mu$m continuum emission and the cometary \hii\ region. 
The H40$\alpha$ line from ATOMS is significantly detected toward the \hii\ region (see panel (c) of Fig.~\ref{fig:H40_alpha}), as indicated in both Fig.~\ref{fig:H40_alpha} and Fig.~\ref{fig:cont_EO}. 
This detection is consistent with the results from the Red MSX Source survey \citep{Lumsden2013}.

In panel (b) of Fig.~\ref{fig:cont_EO}, we present a two-color-composite image of CO\,(2--1) intensity map integrated over the velocity range of \bluevrange and \redvrange. The integrated velocity ranges avoid the channels close to the systemic velocity ($V_{\rm sys}=-48.6$\,\kms), where the CO emission is heavily contaminated by ambient cloud emission. The CO\,(2--1) spectrum is shown in Fig.~\ref{fig:CO_spectrum}.
Explosive outflows are identified in the CO\,(2--1) emission using a two-dimensional dendrogram analysis implemented with \texttt{astrodendro}\footnote{\url{http://www.dendrograms.org}}.
This method characterizes the hierarchical structure of the emission by tracing how isointensity surfaces merge and split as a function of intensity level \citep{Rosolowsky2008}. 
The \texttt{astrodendro} parameters are set to a minimum threshold \texttt{min\_value} of $3\sigma$, \texttt{min\_delta} of $1\sigma$, and a minimum structure size \texttt{min\_pix} corresponding to 1.5 synthesized beam areas.
The identification is carried out on channel maps constructed by integrating the CO emission over velocity bins of 2\,\kms, spanning the range from $-92$ to $+13$\,km\,s$^{-1}$.
To minimize contamination from ambient dense gas, channels close to the systemic velocity ($-54$ to $-37$\,km\,s$^{-1}$) are excluded. The identified leaf structures are shown in Fig.~\ref{fig:channel_map}.

These leaf structures are subsequently linked in position--position--velocity (PPV) space, yielding 29 streamer-like structures. The streamer-like structures are shown by the cyan and red curves in panels (b) and (c) of Fig.\ref{fig:cont_EO} and seem to be radially distributed from a common center.
However, because the automated linking procedure is based primarily on spatial and velocity continuity, the resulting sample may also include unrelated high-velocity structures or streamers associated with local core-driven outflows.
We therefore further examine the individual PV structures and exclude BF1--BF7, RF1, RF2, RF4, RF7, RF8, and RF11 from the candidate explosive-streamer sample.
These streamers show substantial departures from an approximately monotonic increase of $|v-v_{\rm sys}|$ with projected distance and may instead be associated with local core-driven outflows. The final sample contains \numtot candidate explosive-outflow streamers, \numblue of them are blueshifted and \numred of them are redshifted.
The SiO\,(5--4) emission also reveals filamentary structures that partly follow the CO streamers and point towards the inferred outflow origin.

Following the geometric method of \citet{Guzman2024}, we calculate the pairwise intersection points of the streamer trajectories and determine the median as the explosion center located at (RA, DEC)$_{\rm J2000}=(16^{\rm h}15^{\rm m}45.6{\pm}0.2^{\rm s}, -50^{\circ}55'53.9''{\pm}1.04'')$.

\subsection{Physical parameters of outflows}
In this section, we estimate the mass, momentum, energy and dynamical age of the explosive outflows.
Assuming local thermodynamic equilibrium (LTE) and optically thin CO emission, the CO column density can be derived as \citep{Garden1991, Sanhueza2012, Mangum2015}
\begin{equation}
    N_{\mathrm{thin}}(\mathrm{CO})=\frac{8\pi \nu ^3}{c^3}\frac{Q_{\mathrm{rot}}}{g_{u}A_{ul}}\frac{e^{\frac{E_l}{k_{\mathrm{B}}T_{\mathrm{ex}}}}}{1-e^{-\frac{h\nu}{k_{\mathrm{B}}T_{\mathrm{ex}}}}}\frac{\int T_{\mathrm{mb}}dv}{J(T_{\mathrm{ex}}) - J(T_{\mathrm{bg}})}
\end{equation}
and
\begin{equation}
    J(T)=\frac{h\nu}{k_{\mathrm{B}}}\frac{1}{e^{\frac{hv}{k_{\mathrm{B}}T}}-1},
\end{equation}
where $\nu$ is the frequency of the CO\,(2--1) transition ($230.538$\,GHz), $c$ is the speed of light, $Q_{\rm rot}$ is the partition function ($=1/3+\frac{k_{\mathrm{B}}T_{\mathrm{ex}}}{hB}$), $g_{\rm u}$ is the statistical weight of the upper level ($=5$), $A_{\rm ul}$ is the Einstein coefficient for spontaneous emission ($=6.911\times 10^{-7}\,\mathrm{s}^{-1}$), and $E_l$ is the energy of the lower level ($=5.53$\,K). 
$T_{\rm bg}$ is the cosmic microwave background temperature, $T_{\rm mb}$ is the main-beam brightness temperature, and $T_{\rm ex}$ is the excitation temperature. 
We adopt a dust temperature of 24\,K \citep{Urquhart2018, LiuTie2020} as the excitation temperature.

Following \citet{Goldsmith1999}, \citet{Sanhueza2012}, and \citet{Mangum2015}, we correct the CO column density for optical depth using
\begin{equation}
    N(\mathrm{CO})=N_{\mathrm{thin}}(\mathrm{CO})\frac{\tau}{1-e^{-\tau}}.
\end{equation}
Assuming that CO\,(2--1) and $^{13}$CO\,(2--1) share the same excitation temperature, the optical depth $\tau$ can be estimated from
\begin{equation}
    \frac{T_{\mathrm{mb}}(\mathrm{CO})}{T_{\mathrm{mb}}(^{13}\mathrm{CO})}=\frac{1-e^{-\tau}}{1-e^{-\tau /r}},
\end{equation}
where $r$ is the abundance ratio between CO and $^{13}$CO, given by $r=(7.5\pm 1.9)D_{\mathrm{GC}}+(7.6\pm 12.9)$ \citep{Wilson1994}. 
The Galactocentric distance of I16119 is $D_{\rm GC}=5.8$\,kpc \citep{LiuTie2020}.

We then estimate the total mass ($M$), momentum ($P$), energy ($E$), and dynamical age ($t_{\rm dyn}$) of the outflows using
\begin{equation}
    M=d^2 \left[\frac{\mathrm{CO}}{\mathrm{H_2}}\right]^{-1}\mu _{\mathrm{H_2}}m_{\mathrm{H}}\Omega \sum_{i}N_i,
\end{equation}
\begin{equation}
    P=\sum_{i}M_i v_i,
\end{equation}
\begin{equation}
    E=\frac{1}{2}\sum_{i}M_i v_i^2,
\end{equation}
\begin{equation}
    t_{\mathrm{dyn}}=\frac{L_{\rm lobe}}{v_{\rm lobe}},
\end{equation}
where $N_i$ and $v_i$ are the column density and velocity in each velocity channel. 
We adopt a CO abundance of $[\mathrm{CO}/\mathrm{H_2}]=10^{-4}$ \citep{Frerking1982}, a source distance of $d=3.42$\,kpc \citep{LiuXC2024}, and a mean molecular weight per H$_2$ molecule of $\mu_{\rm H_2}=2.8$ \citep{Kauffmann2008}. 
$m_{\rm H}$ is the mass of a hydrogen atom, and $\Omega$ is the solid angle of a single pixel.

The dynamical age is estimated assuming a projected lobe length of 0.24\,pc and a maximum streamer velocity of $\sim40$\,km\,s$^{-1}$. 
Since the observed velocities ($v_i$) represent only the line-of-sight component, we adopt a mean inclination angle of $\theta=57.3^{\circ}$ \citep{Dunham2014, Baug2021, Hoque2026}. 
Accordingly, the momentum and energy are corrected by factors of $1/\cos\theta$ and $1/\cos^2\theta$, respectively.

The derived physical parameters are summarized in Table~\ref{tab:OutflowsParams}.

\begin{table*}
    \centering
    
    \caption{Physical properties of outflows}
    \label{tab:OutflowsParams}
    \begin{tabular}{cccccc}
        \hline\hline
         & Mass &  Momentum & Energy & Dynamical age & References \\
         & \msun & \msun\kms & erg & yr &  \\
        \hline
         I16119-5048 & \MassMsun & \MomentumMsunkms & \EnergyErg & $\sim\tdyn$ & This work\\
         Orion BN/KL & $>8$ & $>160$ & $\sim10^{47-48}$ & $\sim500$ & \citet{Bally2017} \\
         DR21 & 120--210 & $\sim10^{3}$ & $\sim10^{48}$ & $\sim8600$ & \citet{Guzman2024} \\
         G5.89-0.39 & 3.3 & 96 & $\sim10^{46-49}$ & $\sim$1000 & \citet{Klaassen2006, Zapata2020} \\
         S106 & 3.1 & 225 & $>10^{47}$ & $\sim$3500 & \citet{Bally2022} \\
         I16076-5134 & 138--216 & $\sim10^{4}$ & $\sim10^{48-49}$ & $\sim$3500 & \citet{Guzman2022} \\
         I12326-6245 & $\sim$7 & $\sim$800 & $\sim10^{48}$ & $\sim$700 & \citet{Zapata2023} \\
         G34.26+0.15 & $\sim$264 & $4.3\times10^{3}$ & $\sim10^{48}$ & $\sim$19,000 & \citet{Issac2025} \\
         I15520–5234 & 23.8 & 3129.8 & $4.1\times10^{48}$ & $\sim6550$ & \citet{Hoque2026} \\
        \hline
    \end{tabular}

\end{table*}

\subsection{Evidence that I16119 is an explosive outflow candidate}
\label{sect:confirm_explosive}
\begin{figure}
    \centering
    \includegraphics[width = 0.49\textwidth]{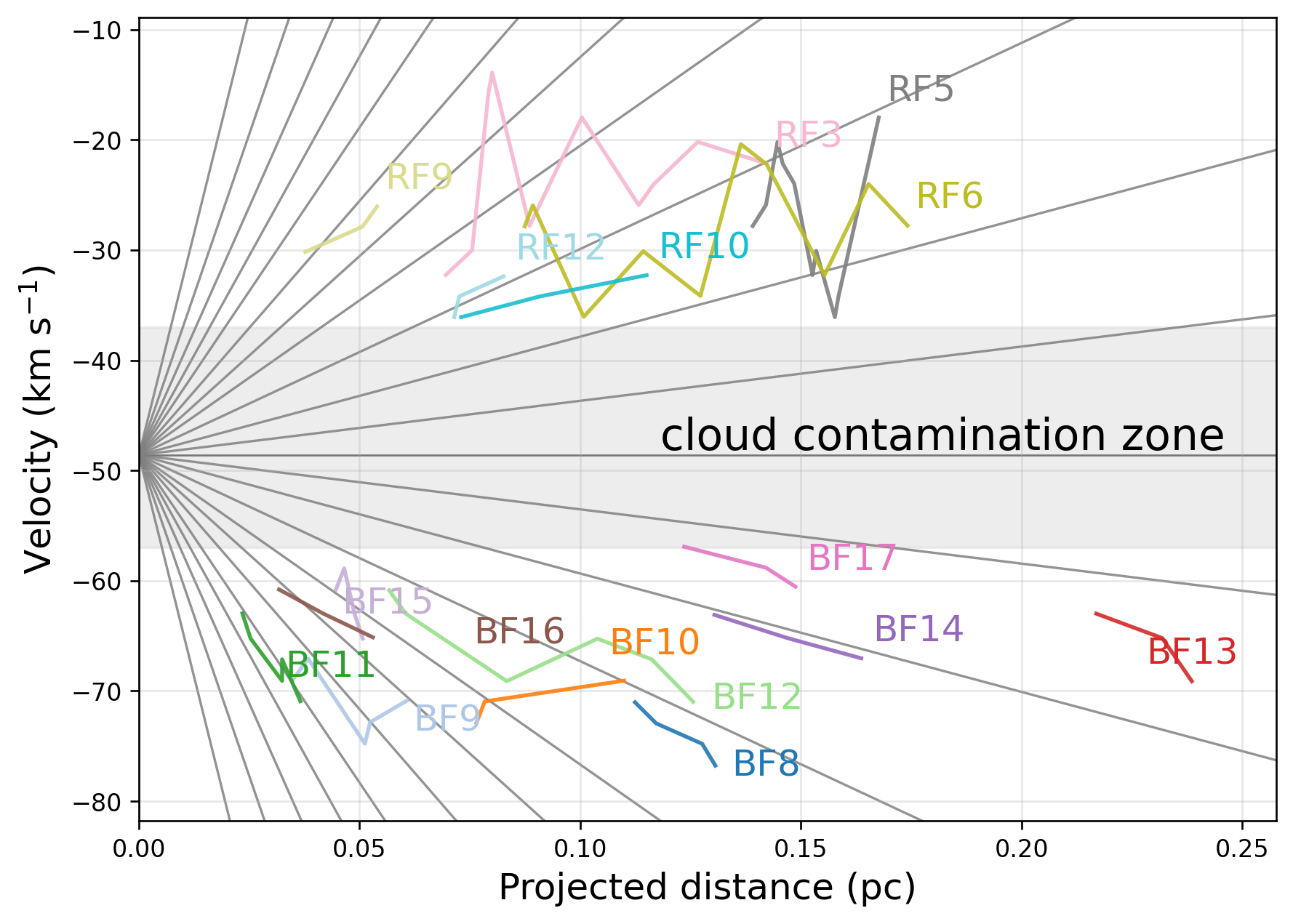}
    \caption{Position-velocity diagram of CO(2-1) streamer structures: radial velocity as a function of projected distance from the common origin for each of the \numtot streamers. Each streamer is represented by a differently colored polyline.. The gray lines depict linear trends between the projected distance and radial velocity, starting from a projected distance of $0''$ and systemic velocity of I16119 at $-48.6$\,\kms. The gray portion indicates the region where the emission is contaminated by the ambient cloud.}
    \label{fig:pv_streamers}
\end{figure}

Explosive outflows are commonly identified through a combination of morphological and kinematic features, including multiple narrow streamers that diverge from a common origin and a Hubble--Lema\^itre-like increase of the line-of-sight velocity with projected distance \citep[e.g.,][]{Zapata2009, Zapata2013, Zapata2020, Zapata2023, Guzman2024, Issac2025, Hoque2026}. In I16119, the CO\,(2--1) streamer trajectories show a broad radial distribution about a compact central region (Fig.~\ref{fig:channel_map}), and several of them exhibit an approximately monotonic increase of $|v-v_{\rm sys}|$ with projected distance in the PV diagram (Fig.~\ref{fig:pv_streamers}). These properties are qualitatively consistent with those expected for an explosive outflow.

Nevertheless, even after excluding the most discrepant streamers, the PV structure is not uniformly described by a single Hubble--Lema\^itre relation. Some of streamers show substantial departures from a simple linear trend, and some features may be affected by unrelated or partially blended core-driven outflows. This is not unexpected given the complexity of I16119, where several dense cores and active star-forming sites are present within the same central region.
Indeed, the CH$_3$OH\,(4$_2$--3$_1$) integrated intensity map shown in Fig.~\ref{fig:outflow_tracers} reveals emission associated with most of the 1.3\,mm dense cores, suggesting that a significant fraction of the core population is already in the protostellar stage.
Local outflow activity driven by these protostellar cores could therefore be superposed on the putative explosive streamers, complicating the observed kinematics and morphology and contributing to deviations from an idealized Hubble--Lema\^itre-like pattern.
In addition, the finite angular resolution and the limitations of the streamer-identification procedure make it difficult to fully separate compact protostellar outflows from putative explosive streamers. Similar caveats have been noted in other explosive outflow systems, where the overall morphology and kinematics favour an explosive interpretation even though individual streamers may deviate from the idealized pattern \citep[e.g.,][]{Issac2025}.

Some streamers also show curvature or flattening in the PV diagram, suggestive of either deceleration or contamination by local gas motions. For example, RF6 may trace material that has interacted with the surrounding molecular gas. In other cases, such as BF10, the streamer paths pass close to dense cores; the observed deviations may therefore reflect interaction with dense core material, or the contribution of outflows launched by those cores. The SiO integrated intensity map provides supporting evidence for shocked gas along parts of the streamer (Fig.~\ref{fig:cont_EO}c), consistent with interactions between high-velocity ejecta and the ambient medium. Such interactions may be enhanced by gas swept up by the externally expanding \hii\ region around I16119, as discussed in Section~\ref{sect:interaction}.

As summarized in Table~\ref{tab:OutflowsParams}, the estimated total outflow mass, momentum, kinetic energy, and dynamical age are \MassMsun\,\msun, \MomentumMsunkms\,\msun\,\kms, \EnergyErg\,erg, and \tdyn\,yr, respectively. These values imply an outflow entrainment rate of $\dot{M}_{\rm out}\sim3\times10^{-4}\,\msun\,{\rm yr}^{-1}$ and a momentum rate of $F_{\rm out}\sim10^{-2}\,\msun\,\kms\,{\rm yr}^{-1}$. Compared with previously reported explosive outflows, the kinetic energy of I16119 is at least one order of magnitude lower, and its total outflow mass is also relatively small. Thus, if I16119 is indeed associated with an explosive event, it would represent a comparatively low-energy member of this class.

However, the inferred outflow activity remains strong when compared with more typical protostellar outflows. For example, \citet{YangAY2018} summarized typical outflow parameters and showed that high-mass outflows generally have $\dot{M}_{\rm out}\sim10^{-5}$--$10^{-3}\,\msun\,{\rm yr}^{-1}$ and $F_{\rm out}\sim10^{-4}$--$10^{-2}\,\msun\,\kms\,{\rm yr}^{-1}$. The values derived for I16119 lie towards the upper end of these ranges. A more relevant comparison can be made with the ALMA study of massive protoclusters by \citet{Baug2021}. Their sample contains 11 protoclusters associated with \hii\ regions; however, two of them, I16076--5134 and I15520--5234, have subsequently been suggested to host explosive outflows \citep{Guzman2022, Hoque2026}. Excluding these two explosive-outflow sources, the total outflow energy derived here for I16119 is higher than those of the remaining sources in the \citet{Baug2021} sample, except I16071--5142. The latter source contains an outflow-driving core with a mass exceeding $100\,\msun$, whereas the dense cores in I16119 typically have masses of only a few solar masses (see Table~\ref{tab:CoresPara}). This suggests that, although I16119 is less energetic than most known explosive outflow systems, its molecular outflow is still unusually powerful relative to typical core-driven outflows.

We therefore interpret I16119 as a plausible explosive outflow candidate. The radial arrangement of the CO streamers, the approximately Hubble--Lema\^itre-like behaviour of several features, the association of SiO emission with the streamer paths, and the relatively large outflow rates compared with typical protostellar outflows together favour an explosive-event scenario. However, the significant deviations from a simple Hubble--Lema\^itre-like velocity pattern, together with possible contamination from core-driven outflows and environmental interactions, indicate that higher angular resolution and sensitivity will be required to establish the explosive nature of the system more robustly.

\subsection{Interaction between 3\,mm clumps and ionized gas}
\label{sect:interaction}

\begin{figure*}
    \centering
    \includegraphics[width = 0.95\textwidth]{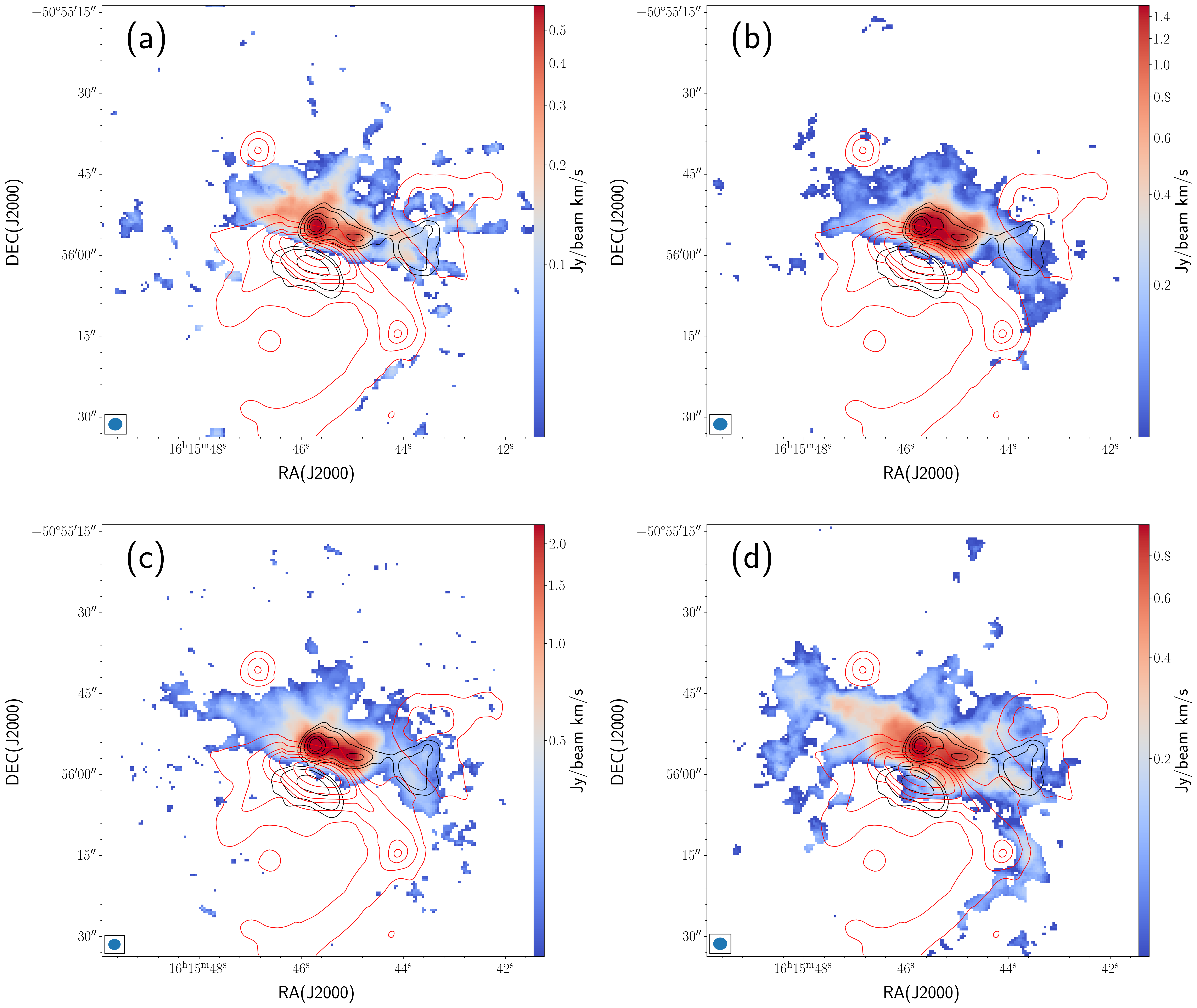}
    \caption{
    Panels (a)–(d) show the integrated intensity maps of \hsco\,(1--0), \hscn\,(1--0), \hcsn\,(11--10), and CCH\,(1--0), respectively. 
    Red contours show the \textit{Spitzer} 8\,$\mu$m emission at [3, 5, 8, 10, 13, 18, 25]$\times\sigma$ ($\sigma=107\,\mathrm{MJy\,sr^{-1}}$). 
    Black contours indicate the ALMA 3\,mm emission at [0.1, 0.15, 0.3, 0.5, 0.7]$\times$ the peak intensity ($0.024\,\mathrm{Jy\,beam^{-1}}$).
    The synthesized beam for each molecular line is shown in the bottom-left corner of the corresponding panel.
    }
    \label{fig:mol}
\end{figure*}

We now examine whether the ionized environment around I16119 plays an active role in shaping the outflow phenomenology. Many well–studied explosive-outflow sources---including Orion BN/KL, DR\,21, G5.89$-$0.39 and G34.26+0.15---are spatially and kinematically linked to nearby or embedded \hii\ regions (e.g., \citealt{Zapata2009, Zapata2013, Zapata2020, Zapata2023, Guzman2024, Issac2025, Hoque2026}). In I16119, the outflow region lies adjacent to a cometary \hii\ region (Fig.~\ref{fig:H40_alpha} and Fig.~\ref{fig:cont_EO}), with H40$\alpha$ emission detected toward the ionized nebula and PDR tracers (e.g., CCH) delineating the UV-illuminated shell (Fig.~\ref{fig:mol}). Together with the anisotropic distribution of streamers and their PV evidence for deceleration, this morphology strongly suggests dynamical coupling between the expanding \hii\ region and the molecular clump that hosts the explosive outflows. Below, we quantify this interaction by comparing the internal (molecular) pressure of the 3\,mm clump with the external (ionized) pressure of the adjacent \hii\ region, and we discuss the implications for the confinement and kinematics of the streamers.

Panels (a), (b), and (c) of Fig.\ref{fig:mol} present the integrated intensity maps of \hsco\,(1--0), \hscn\,(1--0), and \hcsn\,(11--10), respectively. \hsco, \hscn and \hcsn line emissions are thought to be good tracers of high-density gas. In all cases, the dense gas structures form distinct molecular shells adjacent to the \hii region, closely matching the morphology of the 8\,$\mu$m emission.

CCH is thought to be a sensitive tracer for photodissociation regions \citep{Cuadrado2015, Kirsanova2021} and thus effectively marks the boundaries where UV photons actively interact with molecular gas.
Panel (d) of Fig.\ref{fig:mol} is the map of integrated intensity of CCH\,(1--0) line emission overlaid with 8\,$\mu$m and 3\,mm contours. It can be seen that the CCH shell conforms well to the shape of the 8\,$\mu$m continuum emission. The shell structures of molecular line emission and shell structure of 8\,$\mu$m, as well as the adjacent \hii region, indicate the occurrence of feedback from the \hii region on the 3 mm clump where the explosive outflows are located. Besides, the distribution of streamers is notably non-isotropic, in particular, a small number of streamers are detected south-east of the dense gas (Fig.\ref{fig:cont_EO}), which may be attributed to the effect of the adjacent \hii region.
Therefore, we investigate the interaction between the clump containing the explosive outflows and the cometary \hii region.

The molecular gas pressure within the 3\,mm clump can be calculated by
\begin{equation}
    \frac{P_{\rm mol}}{k_{\mathrm{B}}}=n_{\mathrm{env}}T_{\rm eff},
\end{equation}
where $n_{\mathrm{env}}=(1.1\pm0.1)\times 10^{5}\mathrm\,{\rm cm}^{-3}$ is the mean density of the larger-scale 3\,mm clump interacting with the adjacent \hii\ region, estimated from the 3\,mm emission, $k_{\mathrm{B}}$ is the Boltzmann constant, and $T_{\rm eff}$ is the effective kinetic temperature $T_{\rm eff}=\frac{\sigma ^2 _{\rm tot}\mu m_{\rm H}}{k_{\mathrm{B}}}$, where $\mu =2.37$ is the mean molecular weight per free particle \citep{Kauffmann2008} and $m_{\rm H}$ is the mass of the hydrogen atom. $\sigma _{\rm tot}$ is an effective sound speed  including turbulent support derived from
\begin{equation}
    \sigma _{\rm tot}=\sqrt{c_{\rm s}^2 + \sigma _{\rm nt}^2},
\end{equation}
\begin{equation}
    \sigma _{\rm nt}=\sqrt{\sigma _{\rm mol}^2 - \frac{k_{\rm B}T_{\rm k}}{m_{\rm mol}}}
    \label{eq:sigma_nt}
\end{equation}
and
\begin{equation}
    c_{\rm s}=\sqrt{\frac{k_{\mathrm{B}}T_{\rm k}}{\mu m_{\rm H}}},
\end{equation}
where $c _{\rm s}$ is the thermal sound speed, $T_{\rm k}$ is kinetic temperature assumed to be equal to the dust temperature of $\sim24$\,K \citep{Urquhart2018, LiuTie2020}, $m_{\rm mol}$ is the mass of \hsco molecule and $\sigma _{\rm mol}=1.5$\,\kms is the velocity dispersion derived from the \hsco line. The derived $\sigma _{\rm nt}$ and $\sigma _{\rm tot}$ are 1.53\,\kms and 1.55\,\kms, respectively. Therefore, the effective temperature $T_{\rm eff}$ is $\sim692$\,K and molecular gas pressure $P_{\rm mol}$ is $(7.0\pm0.7)\times 10^{7}$\,K\,$\rm cm^{-3}$\,$k_{\rm B}$.

The ionized gas pressure inside the \hii region is the sum of thermal pressure and turbulent pressure, which can be derived from
\begin{equation}
    \frac{P_{\rm i}}{k_{\mathrm{B}}}=2n_{e}T_{e}+n_{e}\mu _{g}m_{\rm H}\sigma _{\rm nt}^{2}/k_{\mathrm{B}},
\end{equation}
\begin{equation}
    n_{e}=\sqrt{\frac{L(\mathrm{H}40\alpha)\nu}{c\epsilon V_{\mathrm{H40\alpha}}}},
\end{equation}
and
\begin{equation}
    L(\mathrm{H}40\alpha)=4\pi d^2 S\Delta v,
\end{equation}
where $\mu _{g}=1.4$ considering both hydrogen and helium gas. $\nu$ is the frequency of H40$\alpha$ line, $\epsilon$ is the efficiency factor ($1.99\times10^{-32}$\,cm$^3$\,erg\,s$^{-1}$), $V_{\mathrm{H40\alpha}}$ is the effective volume of \hii region and $d$ is the source distance. $T_{e}$ is equal to 10000\,K following \citet{Garay1993} and \citet{LiuTie2017}. The $\sigma _{\rm nt}$ is derived from the H40$\alpha$ line of ALMA observations as $\sim8.6$\,\kms. The derived ionized gas pressure $P_{\rm i}$ is $(5.8\pm0.3)\times 10^{7}$\,K\,$\rm cm^{-3}$\,$k_{\rm B}$.

Based on the above calculations, the ionized gas pressure is found to be comparable to the molecular gas pressure within the 3\,mm clump where the candidate explosive outflow is located. This suggests that the expanding \hii\ region could exert a non-negligible dynamical influence on the molecular material, either by counterbalancing the internal pressure of the clump or by modifying its local physical conditions. The spatial correspondence between the cometary \hii\ region, the 3\,mm clumps, and the steep molecular boundary traced by the dense-gas and PDR tracers (Fig.~\ref{fig:mol}) is therefore consistent with a scenario in which the ionized gas has compressed and shaped the molecular shell.

This external feedback may also have affected the apparent distribution of the outflow streamers. The maps of $^{13}$CO\,(2--1), SO\,(6--5), and CH$_3$OH\,(4$_2$--3$_1$) show molecular and shock-related emission concentrated around the central clump, but the identified streamer distribution is not isotropic, with comparatively fewer streamers detected toward the south-eastern side of the explosive center (Fig.~\ref{fig:outflow_tracers}).
Such an asymmetry could arise if the expansion of the adjacent \hii\ region has modified the density structure of the ambient gas, preferentially confining, decelerating, or disrupting streamers propagating in the south-eastern direction.
A similar non-isotropic streamer distribution has been reported in the explosive outflow source G34.26+0.15 \citep{Issac2025}.
Thus, the feedback from the cometary \hii\ region in I16119 may be important not only in shaping the shell-like molecular structure and the physical conditions of the clumps, but also in influencing the observed morphology and kinematics of the candidate explosive outflow.

\subsection{Dense cores revealed by the higher resolution QUARKS 1.3\,mm dust emission}
\label{sect:cores_iden}
\begin{figure}
    \centering
    \includegraphics[width=0.48\textwidth]{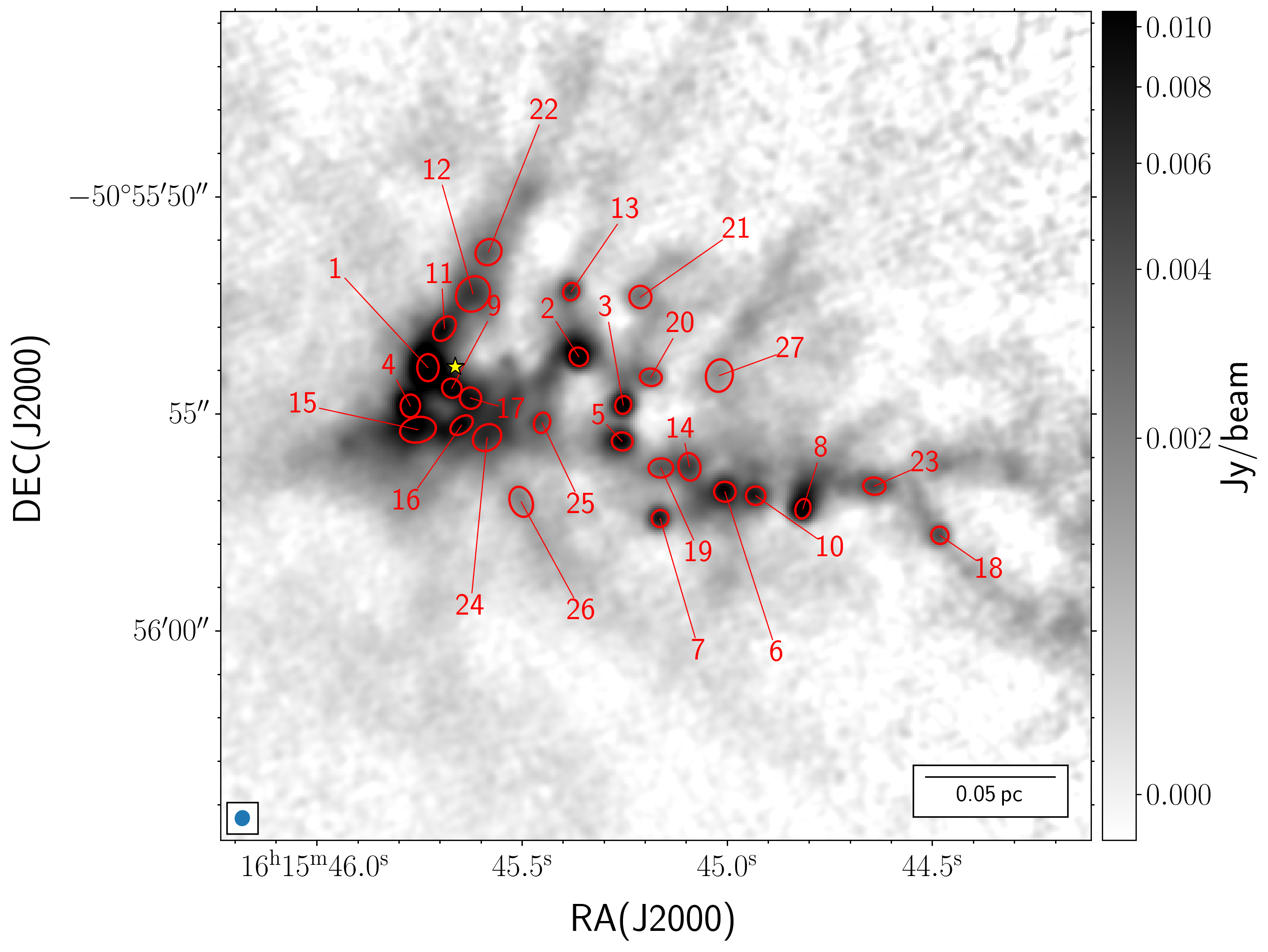}
    \caption{
    Zoomed-in view of 1.3\,mm continuum emission map toward I16119.
    Red ellipses indicate the dense cores identified using the \textit{getsf} source-extraction algorithm, with numbers labeling their corresponding IDs. 
    The yellow star marks the inferred explosion center.
    The scale bar corresponds to 0.05\,pc ($\sim3^{\prime\prime}$) at the adopted distance of 3.42\,kpc.
    The synthesized beam is shown in the lower-left corner.
    }
    \label{fig:dense_cores}
\end{figure}

The QUARKS data offer higher resolution ($\sim$0.3$''$) compared to the ATOMS data, enabling the investigation of smaller-scale structures down to approximately $0.01\,\mathrm{pc}$ at a distance of 3.42\,kpc.
The 1.3\,mm continuum emission from the QUARKS observations is shown in Fig.\ref{fig:dense_cores}, illustrating a highly fragmented filamentary structure extending from the northeast to southwest direction. A total of 27 dense cores are identified from the 1.3\,mm map using the \textit{getsf} algorithm \citep{Menshchikov2021}. Under the assumption that the 1.3\,mm is optically thin, the masses of the dense cores can be derived from 
\begin{equation}
\label{eq:Mass}
    M_{\mathrm{core}} = \frac{R_{\rm gd} S_\nu d^2}{B_\nu(T_{\rm d}) \kappa_\nu},
\end{equation}
where $R_{\rm gd}$ is gas-to-dust mass ratio assumed to be 100, $S_{\nu}$ is the integrated 1.3\,mm flux over the core area, $d$ is the distance of 3.42\,kpc, $B_\nu(T_{\rm d})$ is the Planck function at temperature $T_{\rm dust}$ and the opacity $\kappa_\nu$ is assumed to be 0.899\,cm$^2$\,g$^{-1}$ (MRN model with thin ice mantles, after $10^5$ years of coagulation at gas density of $10^6\,\mathrm{cm}^{-3}$; \citet{Ossenkopf1994}).

After obtaining the core masses, we define the effective core diameter as $D_{\rm eff}=\sqrt{ab}$, where $a$ and $b$ are the major- and minor-axis sizes returned by \textit{getsf}, respectively.
Assuming that the cores are spherical, we adopt an effective radius of $R_{\rm eff}=D_{\rm eff}/2$.
The mean H$_2$ number density is then derived from
\begin{equation}
    n_{\mathrm{H_2}}=\frac{3M_{\mathrm{core}}}{4\pi R_{\mathrm{eff}}^{3}}\frac{1}{\mu_{\mathrm{H_{2}}}m_{\mathrm{H}}},
\end{equation}
where $\mu_{\rm H_{2}}=2.8$ is the molecular weight per H$_{2}$.

The mean H$_{2}$ column densities of dense cores can be calculated using
\begin{equation}
    N_{\mathrm{col}}(\mathrm{H_2})=R_{\mathrm{gd}}\frac{S_{\nu}}{B_{\nu}(T_{\mathrm{d}})\kappa _{\nu}}\frac{1}{\Omega}\frac{1}{\mu_{\mathrm{H_2}}m_{\mathrm{H}}},
\end{equation}
where $\Omega$ is the solid angle of the cores.

Table~\ref{tab:CoresPara} summarizes the physical properties of the identified dense cores.

\begin{table*}
    \centering
    \begin{threeparttable}
    \caption{Physical properties of 1.3\,mm cores}
    \label{tab:CoresPara}
    \begin{tabular}{cccccccc}
        \hline
        Core ID & RA  & DEC  & $S_{\nu}$ & $D_{\rm eff}$ & Mass & $n_{\rm H_{2}}$ & $N_{\rm col}$ \\
         & (J2000) & (J2000) & mJy & au & \msun & $\times 10^6$\,cm$^{-3}$ & $\times 10^{23}$\,cm$^{-2}$ \\
        \hline
        1 & 16h15m45.73s & -50d55m53.92s & 75.2 & 1890 & 15 & 544 & 103 \\
        2 & 16h15m45.36s & -50d55m53.68s & 23.6 & 1460 & 4.8 & 375 & 54.4 \\
        3 & 16h15m45.26s & -50d55m54.79s & 12.8 & 1330 & 2.6 & 268 & 35.5 \\
        4 & 16h15m45.77s & -50d55m54.81s & 25.6 & 1660 & 5.2 & 276 & 45.6 \\
        5 & 16h15m45.26s & -50d55m55.62s & 14.2 & 1530 & 2.9 & 195 & 29.8 \\
        6 & 16h15m45.01s & -50d55m56.78s & 15.1 & 1640 & 3.1 & 168 & 27.5 \\
        7 & 16h15m45.17s & -50d55m57.40s & 7.8 & 1330 & 1.6 & 162 & 21.6 \\
        8 & 16h15m44.82s & -50d55m57.17s & 14.0 & 1350 & 2.8 & 278 & 37.5 \\
        9 & 16h15m45.67s & -50d55m54.40s & 12.6 & 1530 & 2.5 & 173 & 26.3 \\
        10 & 16h15m44.93s & -50d55m56.87s & 7.3 & 1490 & 1.5 & 108 & 16.1 \\
        11 & 16h15m45.69s & -50d55m53.03s & 12.1 & 1800 & 2.5 & 101 & 18.2 \\
        12 & 16h15m45.62s & -50d55m52.23s & 13.8 & 2720 & 2.8 & 33.4 & 9.08 \\
        13 & 16h15m45.38s & -50d55m52.17s & 5.7 & 1320 & 1.2 & 122 & 16 \\
        14 & 16h15m45.09s & -50d55m56.21s & 8.0 & 1940 & 1.6 & 54.3 & 10.5 \\
        15 & 16h15m45.76s & -50d55m55.36s & 18.7 & 2380 & 3.8 & 67.8 & 16.1 \\
        16 & 16h15m45.65s & -50d55m55.25s & 9.8 & 1570 & 2.0 & 124 & 19.4 \\
        17 & 16h15m45.63s & -50d55m54.63s & 8.8 & 1680 & 1.8 & 91.3 & 15.3 \\
        18 & 16h15m44.48s & -50d55m57.79s & 2.5 & 1360 & 0.51 & 48.8 & 6.61 \\
        19 & 16h15m45.16s & -50d55m56.24s & 4.3 & 1710 & 0.87 & 42.3 & 7.2 \\
        20 & 16h15m45.19s & -50d55m54.15s & 3.3 & 1540 & 0.66 & 44 & 6.76 \\
        21 & 16h15m45.21s & -50d55m52.30s & 2.7 & 1750 & 0.54 & 24.4 & 4.26 \\
        22 & 16h15m45.58s & -50d55m51.26s & 4.4 & 2050 & 0.88 & 24.9 & 5.08 \\
        23 & 16h15m44.64s & -50d55m56.65s & 3.1 & 1540 & 0.62 & 41 & 6.3 \\
        24 & 16h15m45.59s & -50d55m55.54s & 6.6 & 2160 & 1.3 & 31.9 & 6.88 \\
        25 & 16h15m45.45s & -50d55m55.20s & 3.1 & 1430 & 0.64 & 53.1 & 7.56 \\
        26 & 16h15m45.50s & -50d55m57.02s & 2.8 & 2070 & 0.57 & 15.5 & 3.2 \\
        27 & 16h15m45.02s & -50d55m54.11s & 2.0 & 2330 & 0.4 & 7.66 & 1.78 \\
        \hline 
    \end{tabular}
    \begin{tablenotes}
        \footnotesize
        \item Notes. Columns (1)--(3) list the core identification IDs and the J2000 coordinates of the core centroid.
        Column (4) gives the integrated 1.3\,mm continuum flux density measured for each core.
        Column (5) gives the effective core diameter, $D_{\rm eff}=\sqrt{ab}$, where $a$ and $b$ are the major- and minor-axis sizes returned by \textit{getsf}, respectively.
        Column (6) gives the core mass derived from the integrated 1.3\,mm flux density.
        Columns (7) and (8) give the mean H$_2$ number density and the mean H$_2$ column density, respectively.
    \end{tablenotes}
    
    \end{threeparttable}
\end{table*}

\subsection{Fragmentation analysis}
\label{sect:Frag}
Since the ALMA 1.3\,mm emission unravels the existence of dense fragments in an explosive outflows region, the properties of fragmentation may be affected by the outflow or the external \hii region. In this section, we investigate the mechanism of fragmentation revealed by the 1.3\,mm continuum emission.

\subsubsection{Collect and collapse process}

The collect and collapse (C\&C) process \citep{Elmegreen1977, Whitworth1994} describes a mechanism in which the expansion of an \hii\ region sweeps up a dense molecular shell. Over time, the accumulated material in this shell may become gravitationally unstable and fragment, potentially giving rise to a new generation of stars. As demonstrated in Section~\ref{sect:interaction}, the molecular clumps are likely influenced by external pressure from the adjacent \hii\ region, motivating an investigation into whether their formation and fragmentation could be explained by the C\&C scenario.

Following the analytical framework in Section~5 of \citet{Whitworth1994}, the characteristic time for the onset of fragmentation ($t_{\rm frag}$), the shell radius at that time ($R_{\rm frag}$), the shell column density ($N_{\rm frag}$), and the typical mass ($M_{\rm frag}$) and separation ($2r_{\rm frag}$) of fragments depend on the initial hydrogen nuclei number density ($n_{\rm ini}$), the sound speed in the pre-shock medium ($c_{\rm s}$), and the ionizing photon rate ($\dot{N}_{\rm ion}$) of the \hii\ region.

We adopt $n_{\rm ini}$ ($\sim$10$^{5}$\,cm$^{-3}$) from ATLASGAL observations \citep{Urquhart2018}, and compute $c_{\rm s}$ assuming a kinetic temperature of 24\,K ($\sim$0.29\,\kms). 
The ionizing photon rate is estimated under the assumption that is equal to the recombination rate, $\dot{N}_{\rm rec}$, and is given by \citep{ZhangC2022, ZhangC2023, ZhangSJ2024}
\begin{equation}
    \dot{N}_{\rm ion} = \dot{N}_{\rm rec} = L_{\rm H40\alpha} \frac{\nu}{c} \frac{\alpha_{\rm B}}{\epsilon}.
\end{equation}
Here $L_{\rm H40\alpha}$ and $\nu$ are the luminosity and frequency of the H40$\alpha$ line, and $\alpha_{\rm B}$ and $\epsilon$ are the total recombination coefficient ($2.54\times 10^{-13}$\,cm$^3$\,s$^{-1}$) and the efficiency factor of H40$\alpha$ photons ($1.99\times 10^{-32}$\,cm$^3$\,erg\,s$^{-1}$), respectively.

Using the adopted values of $n_{\rm ini}$, $c_{\rm s}$, and $\dot{N}_{\rm ion}$ in the analytical model of \citet{Whitworth1994}, we estimate that gravitational fragmentation of the swept-up shell would begin at
$t_{\rm frag}=(0.347\pm0.006)$\,Myr.
At this time, the shell is predicted to have a radius of
$R_{\rm frag}=(0.373\pm0.006)$\,pc and a column density of
$N_{\rm frag}=(3.90\pm0.07)\times10^{22}$\,cm$^{-2}$.
The characteristic fragment mass and separation are predicted to be
$M_{\rm frag}=(15.5\pm0.3)$\,\msun\ and
$2r_{\rm frag}=(0.267\pm0.005)$\,pc, respectively.

To determine whether the \hii region has had sufficient time to trigger such fragmentation, we compare $t_{\rm frag}$ with its dynamical age. Following \citet{ZhangSJ2024}, the dynamical age is estimated as
\begin{equation}
    t_{\mathrm{dyn}}\approx 0.05587\left(\frac{r_{\mathrm{s}}}{\mathrm{pc}}\right)
    \left[\left(\frac{r_{\mathrm{H}40\alpha,\mathrm{eff}}}{r_{\mathrm{s}}}\right)^{7/4}-1\right]\mathrm{Myr},
    \label{eq:tdyn}
\end{equation}
where $r_{\rm s}$ is the initial Str\"omgren radius and $r_{\rm H40\alpha,eff}$ is the current effective radius of the ionized region measured from the H40$\alpha$ emission.
We obtain $t_{\rm dyn}=(1.75\pm0.15)\times10^{-3}$\,Myr, which is approximately 200 times shorter than $t_{\rm frag}$.

This large disparity implies that the swept-up shell has not yet had sufficient time to develop the gravitational instabilities expected in the C\&C scenario. Furthermore, the average separation between dense cores identified in the ALMA 1.3\,mm continuum emission map, as measured using a minimum spanning tree analysis, is considerably smaller than $2r_{\rm frag}$, and the observed core masses are significantly lower than $M_{\rm frag}$ (see Table~\ref{tab:fragmentation}). These inconsistencies in both temporal and physical scales strongly suggest that the fragmentation revealed by the 1.3\,mm emission is unlikely to have been driven by the C\&C process.

This conclusion is consistent with the previous ATOMS study of \citet{ZhangSJ2024}, who found that shell fragmentation is unlikely to occur in early compact \hii\ regions because the expected fragmentation timescale and radius are generally larger than the ages and sizes of the observed \hii\ regions. They further proposed that most dense gas fragments observed in molecular shells of compact \hii region are probably relics of pre-existing Jeans fragmentation in the natal clump, subsequently swept up and reshaped by the expanding \hii\ region, rather than products of the shell fragmentation or C\&C process.

\begin{table}
    \centering
    \begin{threeparttable}
    \caption{Fragmentation at clump-to-core scale}\label{tab:fragmentation}
    \renewcommand{\arraystretch}{1.5}
    \begin{tabular}{lcc}
        \hline
        
         & $M_{\rm core}$ & $\lambda _{\rm core}$ \\ 
         & (\msun) & (pc) \\
        \hline
        
        Observation From 1.3\,mm & 1.6\,\msun & 0.018\,pc \\ 
        
        Thermal Jeans Fragmentation & 0.66\,\msun & 0.020\,pc \\
        
        Turbulent Jeans Fragmentation & 90\,\msun & 0.11\,pc \\
        
        Thermal Cylindrical Fragmentation & 0.54\,\msun & 0.014\,pc\\
        
        Turbulent Cylindrical Fragmentation & 74\,\msun & 0.072\,pc\\
        
        \hline
    \end{tabular}
    \begin{tablenotes}
        \item Note: The core mass (1.6\,\msun) derived from the 1.3\,mm emission is the median mass of all identified dense cores. 
        The observed core separation (0.018\,pc) represents the average separation of dense cores, estimated using the minimum spanning tree (MST) method by dividing the total MST length by $(n-1)$, where $n$ is the number of dense cores.
    \end{tablenotes}
    \end{threeparttable}
\end{table}

\subsubsection{Jeans fragmentation and cylindrical fragmentation}
Following \citet{WangKe2014} and \citet{LiuTie2017}, we calculate the masses and separations of dense cores predicted by Jeans fragmentation using the following equations:
\begin{equation}
    \lambda _{\rm J}=0.066\,\mathrm{pc}\left(\frac{T}{10\,\mathrm{K}}\right)^{1/2}\left(\frac{n_{\mathrm{frag}}}{10^5\,\mathrm{cm}^{-3}}\right)^{-1/2}
\end{equation}
\begin{equation}
    M_{\rm J}=0.877\,\msun\left(\frac{T}{10\,\mathrm{K}}\right)^{3/2}\left(\frac{n_{\mathrm{frag}}}{10^5\,\mathrm{cm}^{-3}}\right)^{-1/2}.
\end{equation}
Here, the particle number density $n_{\rm frag}$ is the mean density of the compact parent clumps in which the 1.3\,mm dense cores are embedded, estimated from ALMA 3\,mm continuum emission ($n_{\mathrm{frag}}\sim2.4\times10^6\,\mathrm{cm}^{-3}$). We note that this density is higher than the mean density adopted for the pressure comparison in Section~\ref{sect:interaction}, because it is measured over the compact parent structure that hosts the 1.3\,mm dense cores, rather than over the larger-scale 3\,mm clump.
The effective temperature $T$ can be calculated using $T=\frac{\mu m_{\rm H}}{k_{\mathbf{B}}}\sigma ^2$. If fragmentation is dominated by thermal motion, $\sigma$ is sound speed $c_{\rm s}$; for the turbulence dominated conditions, $\sigma$ is replaced by $\sigma _{\rm nt}$ estimated from \hsco line (equation \ref{eq:sigma_nt}).

We also examine whether the fragmentation is governed by cylindrical fragmentation, following \citet{WangKe2014}, \citet{LiuTie2017} and \citet{XuFW2024}. The typical spacing of fragmentation and fragment mass can be derived from
\begin{equation}
    \lambda_{\rm cl}=1.24\,\mathrm{pc}\left(\frac{\sigma}{\rm 1\,km\,s^{-1}}\right)\left(\frac{n_{\rm c}}{\rm 10^{5}\,cm^{-3}}\right)^{-1/2}
\end{equation}
and
\begin{equation}
    M_{\rm cl}=575.3\,\msun\left(\frac{\sigma}{\rm 1\,km\,s^{-1}}\right)\left(\frac{n_{\rm c}}{\rm 10^{5}\,cm^{-3}}\right)^{-1/2},
\end{equation}
where $\sigma$ is replaced by sound speed $c_{\rm s}$ if thermal motion dominates, or $\sigma _{\rm nt}$ if turbulence prevails. The central density $n_{\rm c}$ is derived as the average of the 3\,mm clump density and core density, which amounts to $n_{\mathrm{c}}\sim6\times10^{7}\,\mathrm{cm}^{-3}$.

The results are summarized in Table \ref{tab:fragmentation}. The dense cores separations and masses predicted by turbulent Jeans fragmentation and turbulent cylindrical Jeans fragmentation are significantly larger than the observed average separations and masses of dense cores. In contrast, under thermally dominated conditions, separations of observed dense cores ($\sim0.018$\,pc) is comparable to the separation predicted by thermal Jeans fragmentation ($\sim0.020$\,pc) and thermal cylindrical fragmentation ($0.014$\,pc). Although the mean core masses exceed the predictions from thermal Jeans and thermal cylindrical fragmentation, the predicted values are roughly consistent with the masses of half of the dense cores.

Numerous studies have investigated fragmentation in high-mass clumps \citep[e.g.,][]{LiuTie2017, Sanhueza2017, Beuther2018, Sanhueza2019, LuXing2020, ZhangSJ2021, Morii2024, Ishihara2024, ZhangSJ2024}. 
However, a comprehensive analysis of fragmentation is beyond the scope of this work. 
Instead, we compare our results with those of \citet{LiuTie2017}, who studied the G9.62+0.19 complex, a region adjacent to an \hii\ region. 
They found that thermal Jeans fragmentation can reproduce the separations and masses of quiescent cores, but underestimates those of more evolved objects.
Similarly, in I16119, most of the identified dense cores exhibit CH$_3$OH\,(4$_2$–3$_1$) emission (see Fig.~\ref{fig:outflow_tracers}), indicating that a significant fraction of them are already in relatively evolved stages. 
It is therefore plausible that the present core masses have been enhanced by ongoing accretion following their initial fragmentation. 
This interpretation is consistent with the model of \citet{Dib2010mn}, who showed that time-dependent gas accretion can significantly modify the core mass function, producing more massive cores than those expected from the initial fragmentation alone \citep{Dib2010mn}.
Consequently, while the observed core separations are consistent with a thermal Jeans instability, the measured core masses likely reflect subsequent mass growth rather than the initial fragmentation scale.

These results suggest that thermal Jeans mass instability coupled to gas accretion likely dominates the fragmentation and mass growth process of the 3\,mm clumps.

\subsection{Mass Segregation}
\label{sect:mass_segregation}
Mass segregation refers to a spatial distribution in which more massive objects are more centrally concentrated than lower-mass objects, compared with that expected from a random distribution \citep{Allison2009, Parker2015}. 
In dynamically evolved stellar clusters, mass segregation may arise from two-body relaxation, through which massive members lose kinetic energy and sink toward the bottom of the gravitational potential \citep[e.g.][]{Spitzer1969}, or be primordial and an imprint of the star formation process \citep[e.g.][]{Dib2007}. 
Here we examine whether the 1.3\,mm dense cores in I16119 show evidence of mass segregation.

We quantify the degree of mass segregation using the minimum spanning tree (MST) method and the mass segregation ratio $\Lambda_{\rm MSR}$ following \citet{Dib2018}, \citet{Dib2019} and \citet{Xu2024apjs}. 
The dense cores are first sorted in descending order of mass. 
For a given number $N_{\rm MST}$, we construct the MST of the $N_{\rm MST}$ most massive cores and calculate its mean edge length, $l_{\rm massive}$. 
For comparison, we perform 1000 Monte Carlo trials, in each of which $N_{\rm MST}$ cores are randomly selected from the full core sample and the corresponding mean MST edge length, $l_{\rm random}$, is calculated. 
The mass segregation ratio is then defined as
\begin{equation}
    \Lambda_{\rm MSR}(N_{\rm MST}) =
    \frac{\left< l_{\rm random} \right>}{l_{\rm massive}}
    \pm
    \frac{\sigma_{l, \mathrm{random}}}{l_{\rm massive}},
    \label{eq:lambda_msr}
\end{equation}
where $\left< l_{\rm random} \right>$ and $\sigma_{\mathrm{l},\rm random}$ are the mean and standard deviation of the MST lengths obtained from repeated random selections of $N_{\rm MST}$ cores, respectively. 
We further define
\begin{equation}
    f_{\rm MST} = \frac{N_{\rm MST}}{N_{\rm MST,max}},
    \label{eq:f_mst}
\end{equation}
where $N_{\rm MST,max}=27$ is the maximum number of cores considered in the calculation, corresponding to the total number of identified dense cores.
Thus, $f_{\rm MST}$ represents the fraction of cores included in the MST analysis.

By definition, $\Lambda_{\rm MSR}\simeq 1$ indicates that the massive cores are distributed in the same way as the general core population, i.e. no significant mass segregation. 
A value of $\Lambda_{\rm MSR}>1$ means that the MST length of the most massive cores is shorter than expected from random subsets, implying that the massive cores are more spatially concentrated and therefore mass segregated. 
Conversely, $\Lambda_{\rm MSR}<1$ indicates that the massive cores are more widely distributed than the general population, corresponding to inverse mass segregation.

\begin{figure}
    \centering
    \includegraphics[width=0.49\textwidth]{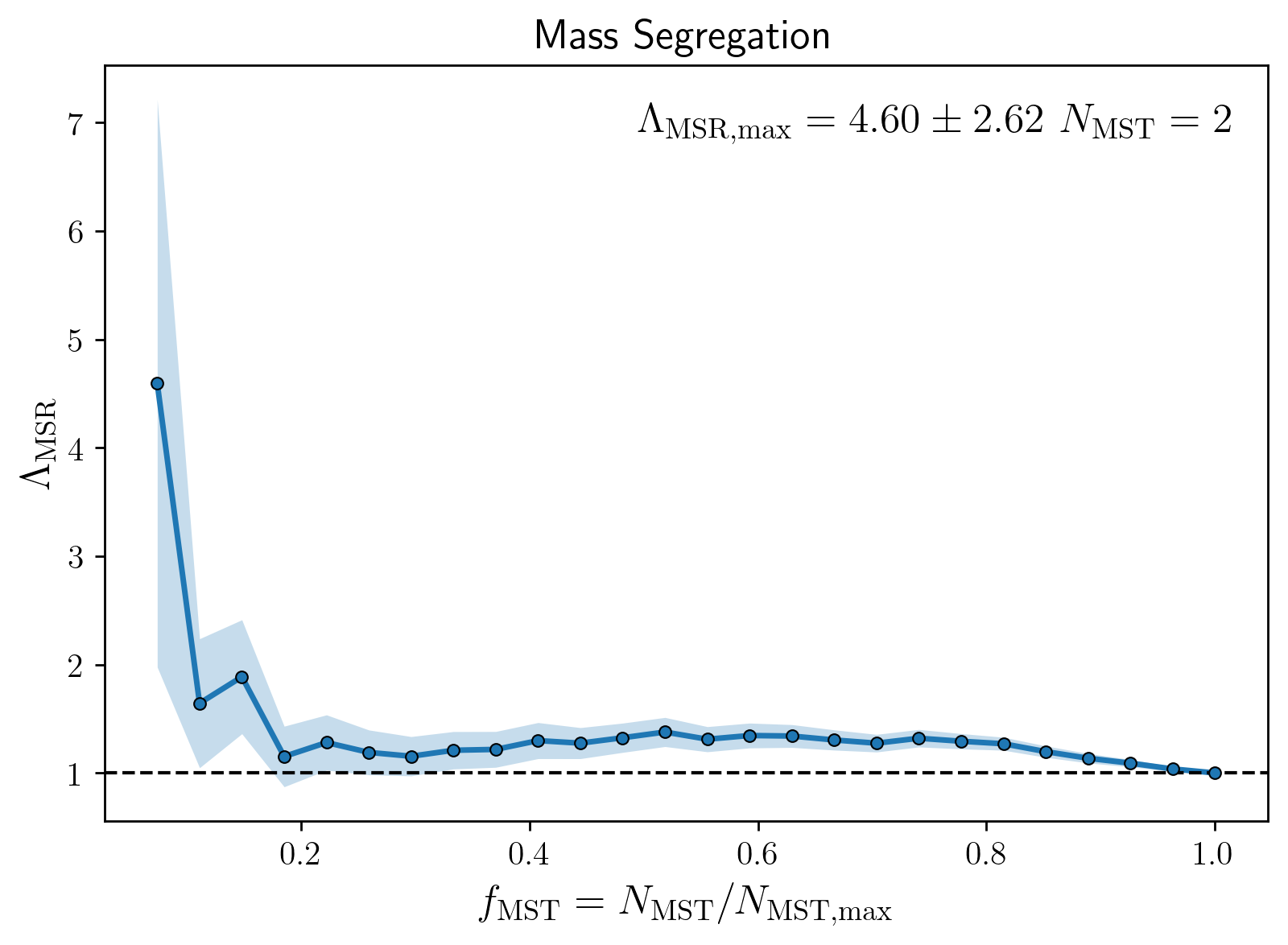}
    \caption{
    Mass segregation ratio, $\Lambda_{\rm MSR}$, as a function of the fraction of selected cores, $f_{\rm MST}=N_{\rm MST}/N_{\rm MST,max}$.
    The blue scatters show the derived $\Lambda_{\rm MSR}$ values, while the shaded region indicates the $1\sigma$ uncertainty.
    The horizontal dashed line marks $\Lambda_{\rm MSR}=1$, corresponding to no significant mass segregation.
    }
    \label{fig:mass_segregation}
\end{figure}

The resulting $\Lambda_{\rm MSR}$ plot for I16119 is shown in Fig.~\ref{fig:mass_segregation}. 
The maximum mass segregation ratio is measured to be 
$\Lambda_{\rm MSR,max} = 4.62 \pm 2.62$ at $N_{\rm MST}=2$. 
This value remains above 1 within the uncertainty, suggesting that the two most massive dense cores are more spatially compact than expected for randomly selected cores. 
As $f_{\rm MST}$ increases, $\Lambda_{\rm MSR}$ gradually decreases towards 1, indicating that the segregation signal is primarily associated with the most massive cores rather than the core population as a whole. 
We therefore find evidence that the 1.3\,mm dense cores in I16119 are mass segregated.

\subsection{Possible Driving Mechanism of Explosive Events}
\label{sect:driving_mechanism}
Several mechanisms have been proposed to explain the origin of explosive outflows. 
In the Orion BN/KL region, the explosive event has been interpreted as the outcome of the dynamical decay of a non-hierarchical massive multiple system, possibly involving close encounters, stellar ejections, or mergers \citep{Zapata2009, Bally2011, Bally2017}. 
In this scenario, the rapid rearrangement of masses in a compact massive system can release gravitational energy impulsively and drive a wide-angle, multi-streamer outflow. 
Similar dynamical interactions among massive cores or protostars have also been discussed as a possible origin of other explosive outflow candidates \citep[e.g.,][]{Hoque2026}.

In I16119, the dense cores show evidence of mass segregation (Section~\ref{sect:mass_segregation}), suggesting that the most massive cores are preferentially concentrated towards the central region. 
Several massive dense cores close to the inferred explosive center, such as MM1, MM4, and MM15, are among the most massive members of the core population. 
Their projected proximity to the candidate explosive center raises the possibility that the central core system has undergone a rearrangement of masses. 
Dynamical interactions or close encounters among these centrally concentrated massive cores may therefore have contributed to the generation of the candidate explosive outflow.

The 1.3\,mm continuum morphology provides another, albeit tentative, clue. 
The dense cores near the center appear to surround a region of relatively weak continuum emission, forming a cavity-like structure in projection (Fig.~\ref{fig:dense_cores}). 
In previously studied explosive outflow systems, the inferred explosion center is often located within, or close to, a cavity or shell-like structure associated with the central cluster. 
In I16119, however, the inferred explosive center does not coincide exactly with the apparent 1.3\,mm continuum cavity. 
This mismatch does not necessarily rule out a dynamical-interaction origin. 
The position of the explosive center is estimated from the back-projection of identified streamers, and may be affected by the complex physical environment, projection effects, and the possible blending of explosive streamers with core-driven outflows. 
In addition, if the explosive event occurred several thousand years ago, subsequent motions of the dense cores could have changed their projected positions relative to the original interaction site.

We therefore regard the rearrangement and dynamical interaction of the centrally concentrated massive cores as a plausible, but not yet confirmed, driving mechanism for the candidate explosive event in I16119. 
Higher angular resolution and sensitivity observations will be required to resolve the innermost core system, separate compact protostellar outflows from the putative explosive streamers, and test whether the dense cores show kinematic signatures expected from a recent dynamical interaction.

\section{SUMMARY}
\label{sect:summary}

We have presented a multiwavelength study of the massive star formation region I16119 using ALMA ATOMS and QUARKS observations, together with archival ATCA radio continuum and \textit{Spitzer} mid-infrared data. The main results are the followings:

\begin{enumerate}

\item The CO\,(2--1) emission reveals a complex system of high-velocity streamer-like structures around the central region of I16119. Using a two-dimensional dendrogram analysis of velocity-channel maps and subsequent linking in PPV space, we identify \numtot streamers, including \numred redshifted and \numblue blueshifted features. These streamers are approximately radially distributed and can be traced back to a common origin at (RA,DEC)$_{\rm J2000} = (16^{\rm h}15^{\rm m}45.6^{\rm s}, -50^{\circ}55^{\prime}53.9^{\prime\prime})$. The SiO\,(5--4) emission shows filamentary shocked-gas structures that partly follow the CO streamers, supporting an association between the high-velocity molecular gas and shocks.

\item Several CO streamers exhibit an approximately monotonic increase of $|v-v_{\rm sys}|$ with projected distance from the inferred origin, similar to the Hubble--Lemaître-like velocity pattern observed in previous explosive outflow candidates. However, the PV structure is not
fully described by a single linear relation. Some streamers show curvature, flattening, or deviations from the idealized explosive pattern, possibly due to interaction with ambient dense gas, projection effects, or contamination from compact core-driven outflows. We therefore regard I16119 as a plausible explosive outflow candidate, rather than a confirmed explosive outflow system.

\item Assuming LTE and applying an opacity correction based on the CO/$^{13}$CO ratio, we estimate a total outflow mass of $M_{\rm out}=\MassMsun\,\msun$, a momentum of $P_{\rm out}=\MomentumMsunkms\,\msun\,\mathrm{km}\,\mathrm{s}^{-1}$, a kinetic energy of $E_{\rm out}=$\,\EnergyErg\,erg, and a dynamical age of $t_{\rm dyn}\sim\tdyn$\,yr. The kinetic energy is at least one order of magnitude lower than those of most known explosive outflows, suggesting that I16119, if explosive in origin, represents a relatively low-energy member of this class. Nevertheless, its mass entrainment rate and momentum rate are high compared with typical protostellar outflows in massive star formation regions.

\item The candidate explosive outflow is located adjacent to a cometary \hii\ region traced by ATCA 6\,cm continuum emission and ALMA H40$\alpha$ emission. Dense-gas tracers, including H$^{13}$CO$^{+}$\,(1--0), H$^{13}$CN\,(1--0), and HC$_3$N\,(11--10), as well as the PDR tracer CCH\,(1--0), reveal shell-like structures closely associated with the 8\,$\mu$m emission. The ionized gas pressure is found to be comparable to the molecular gas pressure within the 3\,mm clump, suggesting that feedback from the adjacent \hii\ region can exert a non-negligible dynamical influence on the molecular material. Such feedback may have compressed and shaped the molecular shell, and may also contribute to the asymmetric distribution and possible deceleration of the outflow streamers.

\item The high-resolution ALMA 1.3\,mm continuum emission resolves the central molecular structure into 27 dense cores identified with \textit{getsf}. The dense cores are distributed along a fragmented filamentary structure close to the inferred explosive center. A comparison with the C\&C model shows that the fragmentation time-scale and characteristic fragment separation predicted for an expanding \hii\ region are much larger than the observed values, indicating that the present dense-core population is unlikely to have formed through the C\&C process.

\item The observed mean core separation, estimated from a minimum spanning tree analysis, is $\sim0.018$\,pc, comparable to the values predicted by thermal Jeans fragmentation and thermal cylindrical fragmentation. Although the observed core masses are higher than the purely thermal predictions, this discrepancy may be explained by subsequent mass growth through accretion. We therefore suggest that the initial fragmentation in the 3\,mm clump was likely governed mainly by thermal instability, followed by continued accretion.

\item The 1.3\,mm dense cores show evidence of mass segregation. The maximum mass segregation ratio is $\Lambda_{\rm MSR,max}=4.62\pm2.62$ at $N_{\rm MST}=2$, indicating that the two most massive dense cores are more spatially concentrated than expected from a random distribution. Several of the most massive cores are located close to the inferred explosive center, suggesting that dynamical interactions or close encounters among centrally concentrated massive cores may have contributed to the candidate explosive event.

\end{enumerate}

Overall, the radial arrangement of the CO streamers, the approximately Hubble--Lemaître-like behavior of several features, the association with SiO shocked gas, and the unusually strong outflow activity compared with typical protostellar outflows support the interpretation of I16119 as a low-energy explosive outflow candidate. At the same time, the deviations from an ideal explosive velocity field, together with possible contamination from core-driven outflows and the influence of the adjacent \hii\ region, indicate that higher angular resolution and higher sensitivity observations will be required to confirm the explosive nature of this system and to identify its driving mechanism.

\section*{Acknowledgements}

This work was mainly supported by the National Key R\&D Program of China under grant No.2022YFA1603103 and 2023YFA1608002, the National Natural Science Foundation of China (NSFC) under grant No.12373029, and the Central Guidance for Local Science and Technology Development Fund ZYYD2025ZY23.
It was also partially supported by the Tianshan Talent Training Program under grant No. 2024TSYCTD0013, the NSFC under grant Nos. 12173075, 12103082, and 12403033, the Chinese Academy of Sciences (CAS) “Light of West China” Program under grant No. xbzg-zdsys-202212, and the Natural Science Foundation of Xinjiang Uygur Autonomous Region under grant No. 2023D01A11.
GW acknowledges the support from the Tianchi Talent Program and Natural Science Fund for Distinguished Young Scholars of Xinjiang Uygur Autonomous Region.
MJ acknowledges the support of the Research Council of Finland Grant No. 348342. G.G and L.B. gratefully acknowledge support by the ANID BASAL project FB210003. PS was partially supported by a Grant-in-Aid for Scientific Research (KAKENHI Number JP26H02066, JP23H01221, and JP26K00748) of JSPS. C.W.L is supported  by the Korea Astronomy and Space Science Institute grant funded by the Korea government (MSIT; project No. 2025-1-841-02). Moreover, this research has been funded by the Science Committee of the Ministry of Science and Higher Education of the Republic of Kazakhstan (Grant No. AP26102915).

This paper makes use of the following ALMA data: ADS/JAO.ALMA\#2019.1.00685.S and 2021.1.00095.S. ALMA is a partnership of ESO (representing its member states), NSF (USA), and NINS (Japan), together with NRC (Canada), MOST and ASIAA (Taiwan), and KASI (Republic of Korea), in cooperation with the Republic of Chile. The Joint ALMA Observatory is operated by ESO, AUI/NRAO, and NAOJ.

\section*{DATA AVAILABILITY}

The data underlying this article are available in the ALMA archive.

\bibliographystyle{mnras}
\bibliography{References.bib}

\appendix

\section{Spectrum of CO\,(2-1)}
\begin{figure}
    \centering
    \includegraphics[width=0.48\textwidth]{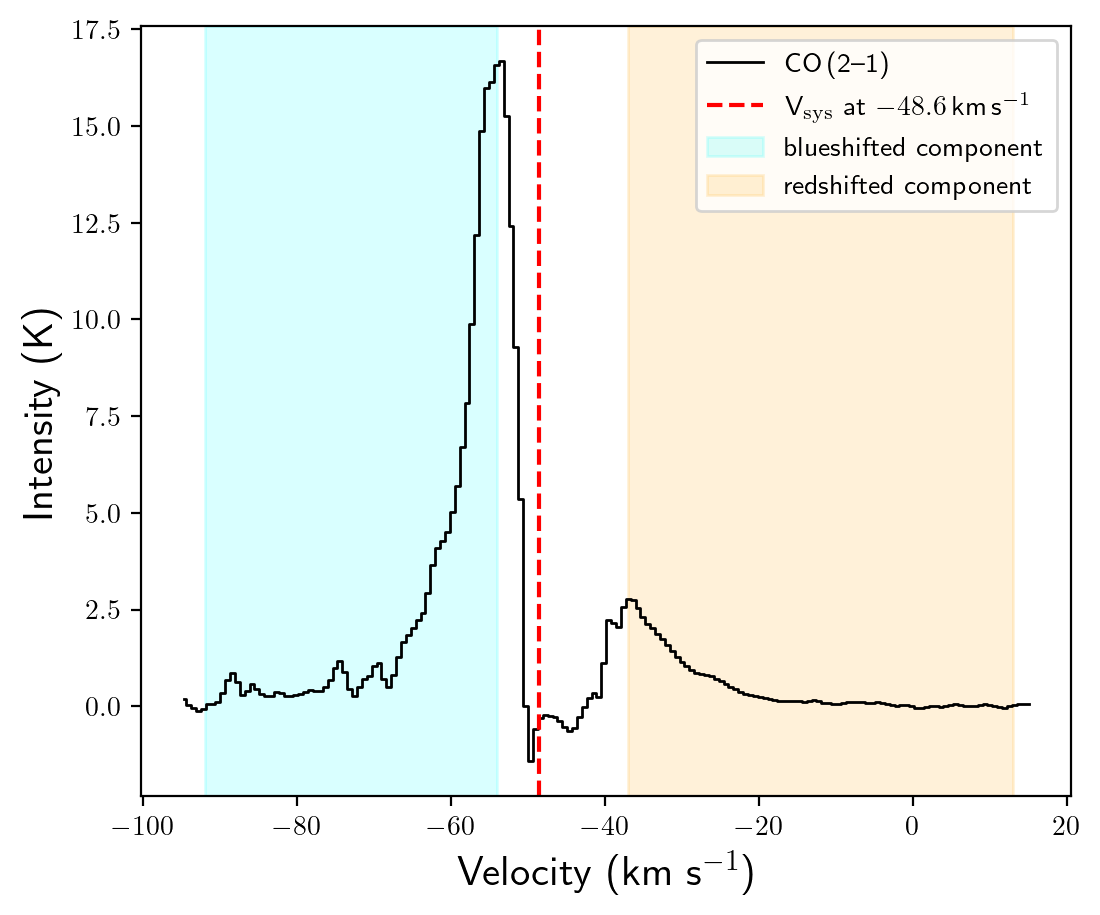}
    \caption{Averaged CO\,(2--1) spectrum extracted within a circular region of diameter $15''$ covering the main explosive outflow area. 
    The vertical dashed red line marks the systemic velocity ($V_{\rm sys} = -48.6$\,km\,s$^{-1}$). 
    The blueshifted ($-92$ to $-54$\,km\,s$^{-1}$) and redshifted ($-37$ to $+13$\,km\,s$^{-1}$) velocity ranges used in this work are highlighted by the cyan and orange shaded regions, respectively.}
    \label{fig:CO_spectrum}
\end{figure}

The CO\,(2--1) spectrum is shown in Fig.~\ref{fig:CO_spectrum}. 
The velocity ranges adopted to define the blueshifted and redshifted outflow components are indicated in the figure.

\section{Channel maps of CO\,(2-1)}
\begin{figure*}
    \centering
    \includegraphics[width=0.75\textwidth]{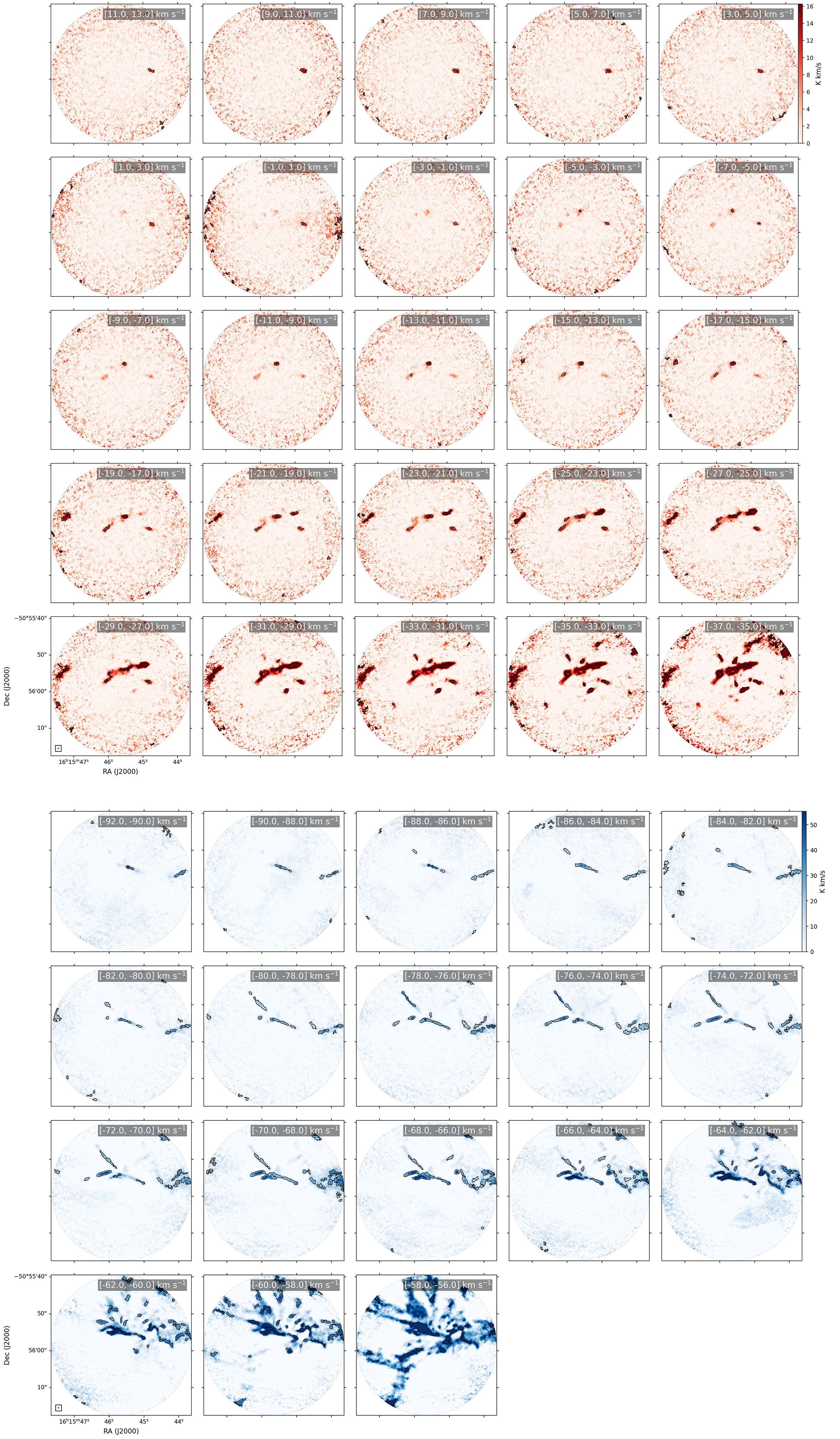}
    \caption{Channel maps of ALMA CO\,(2--1) emission for the redshifted (upper; $V_{\rm lsr} = -37$ to $+13$\,km\,s$^{-1}$) and blueshifted (bottom; $V_{\rm lsr} = -92$ to $-56$\,km\,s$^{-1}$) velocity ranges. The leaf structures identified with the \texttt{astrodendro} algorithm are outlined by black contours.
    The beam of the CO\,(2--1) data is shown in the lower-left corner of the bottom-left panel in each set of channel maps.
    }
    \label{fig:channel_map}
\end{figure*}

The CO\,(2–1) channel maps of width 2\,\kms in the velocity range $-37$ to $+13$\,\kms and $-96$ to $-57$\,\kms, as well as the identified leaf structures, are shown in Fig.\ref{fig:channel_map}.

\section{Moment maps of outflow$/$shocked gas tracers $^{13}$CO, SO and CH$_3$OH\,(4$_2$--3$_1$)}

\begin{figure*}
    \centering
    \includegraphics[width=0.99\textwidth]{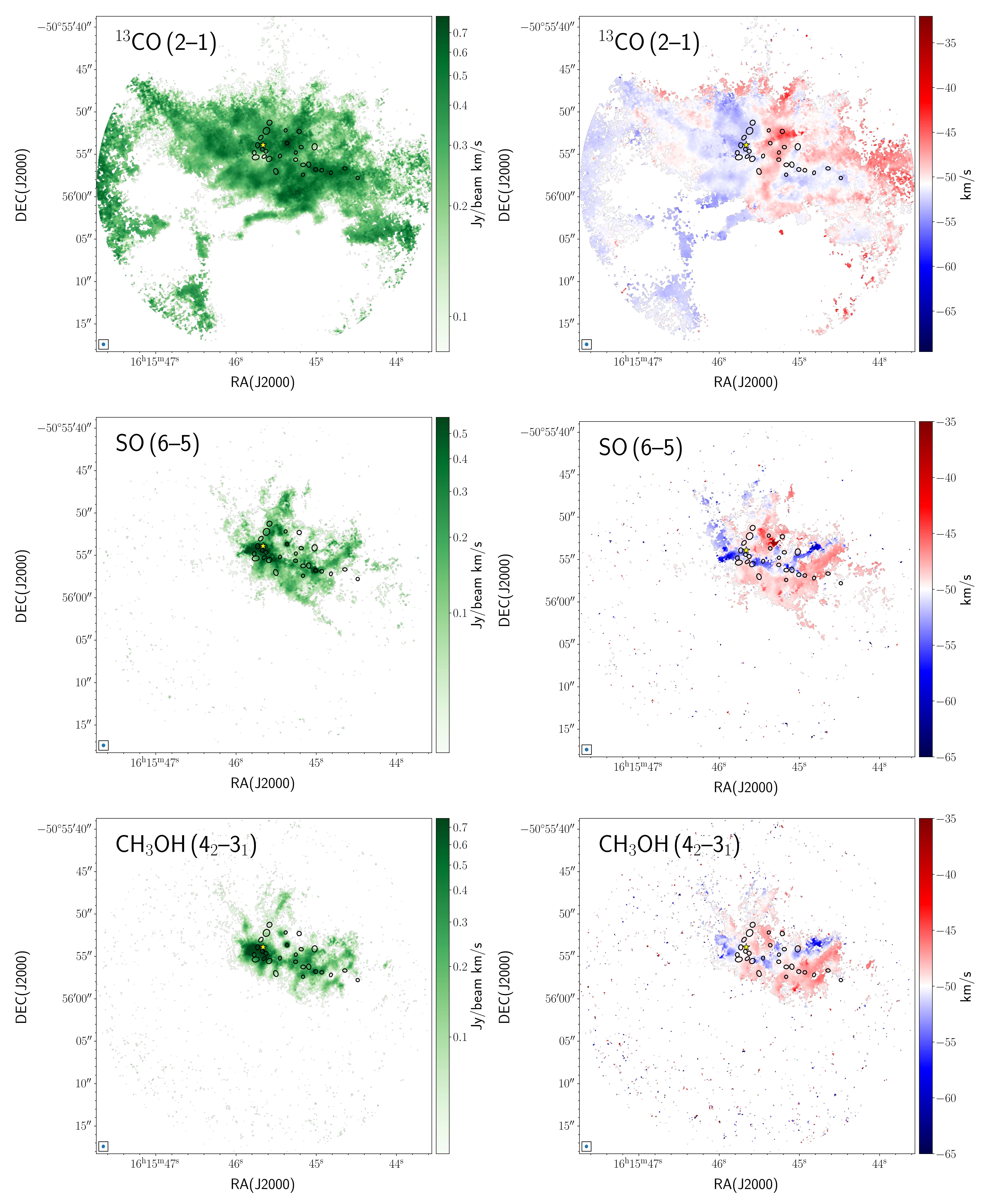}
    \caption{Moment\,0 (left panels) and moment\,1 (right panels) maps of the outflow$/$shocked gas tracers $^{13}$CO\,(2--1), SO\,(6--5) and CH$_3$OH\,(4$_2$--3$_1$). Black ellipses indicate the 1.3\,mm continuum cores. The beam of each molecular line is shown in the lower-left corner of the corresponding panel.}
    \label{fig:outflow_tracers}
\end{figure*}

In this section, we present the moment-0 (integrated intensity) and moment-1 (intensity-weighted velocity) maps of $^{13}$CO\,(2--1), SO\,(6--5) and CH$_3$OH\,(4$_2$--3$_1$) lines obtained from the QUARKS survey. 
Figure~\ref{fig:outflow_tracers} shows the integrated intensity (left panels) maps and velocity (right panels) maps of these molecular tracers. 
The moment maps reveal the morphology and kinematic structure of the outflows, highlighting their spatial distribution and anisotropy.

\bsp	
\label{lastpage}
\clearpage
\noindent
$^{1}$State Key Laboratory of Radio Astronomy and Technology, Xinjiang Astronomical Observatory, CAS, 150 Science 1-Street, Urumqi, Xinjiang, 830011, P. R. China\\
$^{2}$University of Chinese Academy of Sciences, Beijing 100080, PR China\\
$^{3}$Xinjiang Key Laboratory of Radio Astrophysics, Urumqi 830011, PR China\\
$^{4}$State Key Laboratory of Radio Astronomy and Technology, Shanghai Astronomical Observatory, Chinese Academy of Sciences, 80 Nandan Road, Shanghai 200030, People’s Republic of China\\
$^{5}$Departamento de Astronom\'{i}a, Universidad de Chile, Las Condes, 7591245 Santiago, Chile\\
$^{6}$Max Planck Institute for Astronomy, K\"onigstuhl 17, 69117 Heidelberg, Germany\\
$^{7}$Universität Heidelberg, Zentrum für Astronomie, Institut für Theoretische Astrophysik, Albert-Ueberle-Str. 2, 69120 Heidelberg, Germany\\
$^{8}$Rosseland Centre for Solar Physics, University of Oslo, PO Box 1029 Blindern, 0315 Oslo, Norway\\
$^{9}$Institute of Theoretical Astrophysics, University of Oslo, PO Box 1029 Blindern, 0315 Oslo, Norway\\
$^{10}$Department of Physics, University of Helsinki, P.O. Box 64, 00014 Helsinki, Finland\\
$^{11}$Department of Astronomy, School of Science, The University of Tokyo, 7-3-1 Hongo, Bunkyo-ku, Tokyo 113-0033, Japan\\
$^{12}$Institute for Advanced Study, Kyushu University, Fukuoka 819-0395, Japan\\
$^{13}$Department of Earth and Planetary Sciences, Faculty of Science, Kyushu University, Nishi-ku, Fukuoka 819-0395, Japan\\
$^{14}$Korea Astronomy and Space Science Institute, 776 Daedeokdae-ro, Yuseong-gu, Daejeon 34055, Republic of Korea\\
$^{15}$University of Science and Technology, Korea (UST), 217 Gajeong-ro, Yuseong-gu, Daejeon 34113, Republic of Korea\\
$^{16}$Chinese Academy of Sciences South America Center for Astronomy, National Astronomical Observatories, CAS, Beijing 100101, People’s Republic of China\\
$^{17}$S. N. Bose National Centre for Basic Sciences, Block-JD, Sector-III, Salt Lake City, Kolkata 700106, India\\
$^{18}$Institute of Physics and Astronomy, E\"otv\"os Lor\'and University, P\'azm\'any P\'eter s\'et\'any 1/A, H-1117 Budapest, Hungary\\
$^{19}$Faculty of Science and Technology, University of Debrecen, H-4032 Debrecen, Hungary\\
$^{20}$Energetic Cosmos Laboratory, Nazarbayev University, Astana 010000, Kazakhstan\\
$^{21}$Institute of Experimental and Theoretical Physics, Al-Farabi Kazakh National University, Almaty 050040, Kazakhstan
\end{document}